\documentclass{aa}  
\newcommand{\degc}{^{\circ}}

\usepackage{graphicx}
\usepackage{txfonts}
\usepackage{ulem}
\usepackage{amsmath}
\usepackage{booktabs}
\usepackage{lastpage}
\usepackage[dvipsnames]{xcolor}
\usepackage{tabularx}
\usepackage[colorlinks=true, linkcolor=blue, citecolor=blue]{hyperref}

\usepackage{natbib}
\bibpunct{(}{)}{;}{a}{}{,}
\begin{document}

   \title{Probing dust properties through polarized scattered-light images of a sample of ring-shaped protoplanetary disks}
   \titlerunning{Probing dust properties through SPFs}
   
   \author{M. Roumesy\inst{1}
          \and
          F. Ménard\inst{1}
          \and
          G. Duchêne\inst{1, 2}
          \and
          R. Tazaki\inst{3}
          \and
          C. Ginski\inst{4}
          }

   \institute{Univ. Grenoble Alpes, CNRS, IPAG, 38000 Grenoble, France
   \and Astronomy Department, University of California Berkeley, Berkeley CA 94720-3411, USA
   \and Department of Earth Science and Astronomy, The University of Tokyo, Tokyo 153-8902, Japan
   \and School of Natural Sciences, Center for Astronomy, University of Galway, Galway H91 CF50, Ireland
   }
   \date{Received --; accepted --}

  \abstract
  {The evolution of protoplanetary disks, especially in the early stages of planetary formation, as dust grows, is the cornerstone of the birth of planets. The mechanisms involved in the growth of sub-micrometric dust grains into planetesimals within a very short time frame are a challenging field of study, while the initial conditions remain relatively undefined.}
   {One of the main challenges is to unambiguously identify the dust properties within the disk, and our goal is to break this barrier by investigating the light scattered by dust particles lying on the protoplanetary disk surface from many recent promising observations.}
   {In this study, we used a set of 30 polarized light images composed of new VLT/SPHERE observations to examine the light scattered by dust grains. For each ring-shaped system, we used the new {\tt{DRAGyS}} tool to estimate the disk geometry using the substructures visible on the surface and to extract the limb-brightening-corrected scattering phase function, which encodes the dust grains' physical properties. Finally, we compared our results with the {\tt{AggScatVIR}} database of numerical scattering phase functions of nonspherical dust.}
   {We combined our measurement of disk geometry to estimate an average disk flaring of about 1.357. We note some general trends of dust populations in our results. First, we recovered the two categories of scattering phase functions based on their shape, as determined in previous studies. Category I is monotonically decreasing and can be explained by fractal organic aggregates with small monomers of 100nm, or compact aggregates with medium porosity and big monomers of 400nm. Category II is defined by a bell-shaped scattering phase function and can be explained by sub-micrometric irregular grains or compact aggregates with low porosity. This statistical study offers general trends about dust populations, but the degeneracy is too strong to apply this method to a unique disk analysis.
   }
   {
   The extracted scattering phase function from the protoplanetary disk surface is a promising measurement for constraining the properties of dust within the disk. However, their interpretation should be combined with other observations. Multiwavelength investigation, i.e., combining SPFs with disk color measurements, is a promising option.
   }

   \maketitle   
%

\begin{figure*}[ht]
    \centering
    \includegraphics[width=\textwidth]{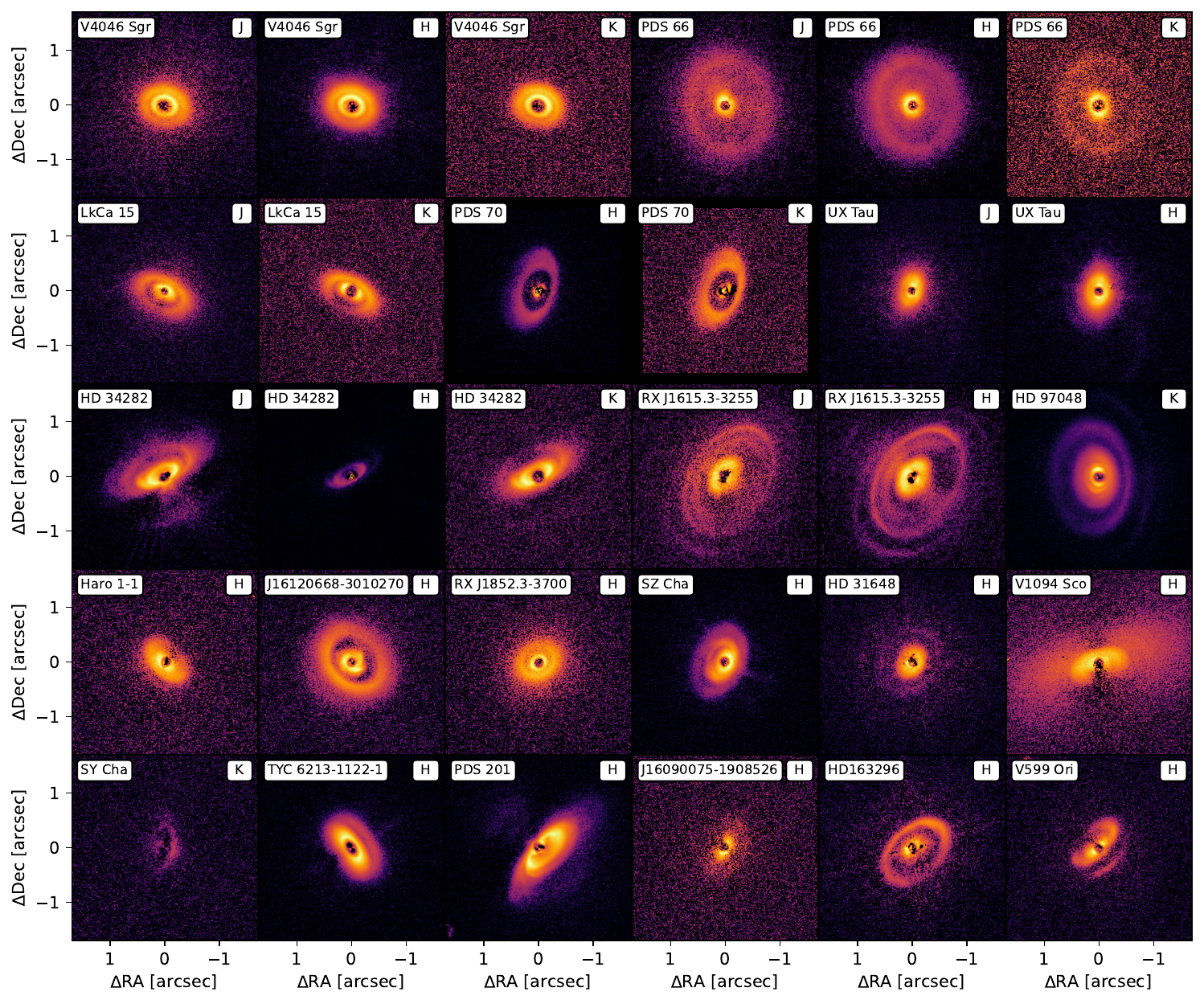}
    \caption{VLT/SPHERE $Q_{\phi}$ observations in the J-, H-, and K-bands. All images have dimensions of $3.4^"\times3.4^"$ and are shown using an individual log-scale color map.}
    \label{FIG:Dataset}
\end{figure*}

\section{Introduction}
Recent observations provide significant insight into the early stages of the planetary formation process, especially through the direct detection of planets in protoplanetary disks, as illustrated by the direct imaging of PDS~70 \citep{keppler_discovery_2018, blakely_james_2025}. These discoveries further confirm that protoplanetary disks, rich in gas and dust, are the sites of ongoing planetary formation \citep{pinte_kinematic_2018, teague_kinematical_2018}. However, the growth of sub-micrometric grains into kilometer-sized planetesimals over a very short timescale remains a mystery in the process of planet formation \citep{drazkowska_planet_2022}. Several studies have investigated the physical conditions and mechanisms governing this process, including the dynamical aspects of dust vertical settling \citep{dubrulle_dust_1995, villenave_introduction_2025}, radial drift \citep{weidenschilling_aerodynamics_1977}, streaming instabilities \citep{youdin_streaming_2005, johansen_protoplanetary_2007, villenave_introduction_2025}, and the influence of dust aerodynamic behavior and coagulation efficiency \citep{ormel_closed-form_2007}.

Today, we are in a position to better understand the physical properties of dust grains evolving in protoplanetary disks, such as their size, shape, and composition, to carry out these investigations and move toward a deeper understanding of the dust growth process. In this context, the number of planet-forming disks observed in scattered light has increased substantially over the past decade, driven by large-scale observational programs such as GTO \citep{beuzit_sphere_2019} and DESTINYS \citep{ginski_disk_2021}. These VLT/SPHERE observations revealed a wide diversity of substructures in the disks, including rings, gaps, and spirals \citep{benisty_optical_2022}, possibly indicating close interactions between the disks and the forming planets. These near-infrared scattered-light observations are strongly dependent on the properties of dust lying on the disk's surface layer \citep{mulders_planet_2013}. It is therefore possible to analyze the ability of the disk's dust to scatter the stellar radiation. This measurement of the polarized or total brightness is the scattering phase function (SPF; \citealt{bohren_absorption_1983}) and is defined by the properties of dust grains \citep{min_multiwavelength_2016, tazaki_light_2016}. The SPF is extracted by measuring the brightness coming from different zones on the disk's surface as a function of the scattering angle. The extraction of the SPF has been frequently performed on debris disks \citep{olofsson_challenge_2020, milli_optical_2019, monnier_polarized_2017}, but its application to protoplanetary disks is relatively new due to their optically thick nature and their more complex geometry caused by the surface flaring. Research on young planet-forming disks started with specific cases such as IM~Lup \citep{tazaki_fractal_2023} to estimate dust properties. Meanwhile, \cite{ginski_observed_2023} analyzed SPFs of polarized light on several disks to study their shape and try to identify general trends. 

It is now established that dust grains are not perfectly spherical and that the Mie theory does not provide the most suitable method for describing their optical properties. Recent work has shown that the grains observed in comets such as 67P/Churyumov-Gerasimenko are fractal aggregates \citep{blum_evidence_2017, mannel_dust_2019, blum_formation_2022}. These results are in line with numerical simulations \citep{lombart_grain_2021, lombart_fragmentation_2022} and laboratory experiments \citep{blum_dust_2018}, in which dust growth by coagulation and fragmentation produces complex aggregates. In this context, the development of the {\tt{AggScatVIR}}\footnote{https://github.com/rtazaki1205/AggScatVIR} database \citep{tazaki_size_2022, tazaki_fractal_2023} represents a significant step forward, providing SPFs for irregular grains and aggregate particles---be they compact or fractal---composed of sub-micrometric monomers.

The complex geometry of protoplanetary disks has long limited statistical studies of SPFs. The new {\tt{DRAGyS}}\footnote{https://github.com/mroumesy/DRAGyS} tool \citep{roumesy_dragys_2025} overcomes this barrier by deriving disk geometry from ring structures and extracting the SPF directly from the disk's surface. This approach generalizes previous individual geometric analyses \citep{de_boer_possible_2021, avenhaus_disks_2018, ginski_direct_2016} and removes the need for time‑consuming radiative-transfer simulations.

In Sect. \ref{SECT:Observations} we present the 30 ring-shaped disk-polarized-intensity $Q_{\phi}$ images observed with the VLT/SPHERE instrument \citep{beuzit_sphere_2019}. Section \ref{SECT:Methods} describes the methods and main results. We first estimate the disk geometry, including flaring in Sect. \ref{SECT:Geometry}, then we describe the SPF extraction method using {\tt{DRAGyS}} in Sect. \ref{SECT:SPF}. Finally, we analyze the dust properties with {\tt{AggScatVIR}} in Sect. \ref{SECT:Dust} to identify general trends in the properties of dust within protoplanetary disks. A discussion of these results is presented in Sect. \ref{SECT:Discussion}.

\section{Observations} \label{SECT:Observations}
The objects selected in this study are moderately inclined protoplanetary disks with one or multiple well-resolved rings in polarized scattered light in the near-infrared bands. Our dataset combines observations from the SPHERE/ESO archives with more recent observations from the Disk Evolution Study Through Imaging of Nearby Stars (DESTINYS) ESO large program \citep{ginski_disk_2020, ginski_disk_2021}, mostly located in the Orion \citep{valegard_disk_2024}, Chamaeleon I \citep{ginski_sphere_2024}, and Taurus \citep{garufi_sphere_2024} star-forming regions. All of these data were obtained using the {\tt{IRDIS}} \citep{dohlen_infra-red_2008} instrument of VLT/SPHERE \citep{beuzit_sphere_2019} in dual-beam polarization mode to obtain $Q_{\phi}$ polarized images with a coronagraph to improve the contrast. The $Q_{\phi}$ images of selected objects are shown in Fig. \ref{FIG:Dataset}.

As an additional source, the disks around TYC~6213-1122-1 and Haro 1-1, both coming from the Scorpius star-forming region, were observed in polarimetry for the first time during the DESTINYS ESO large program (PI:Ginski). We provide an estimate of inclination and position angle for these two systems based on their ring structure using the {\tt{DRAGyS}} tool \citep{roumesy_dragys_2025}. Haro 1-1 lies at $151,1$~pc \citep{gaia_collaboration_vizier_2020}. The disk has an inclination of $54\degc$, a position angle of $225.0\degc \pm 3.8\degc$, and seems to have an inner ring that is partially hidden by the coronagraph. Another ring is detectable, centered around $50$~au from the central star and reaching out to $\sim80$~au. TYC 6213-1122-1 lies at 154.6 pc \citep{gaia_collaboration_vizier_2020}. Its inclination is around $57\degc$, with a position angle of $31.8\degc \pm 4.1\degc$. It reveals a relatively broad ring extending from about $30$~au to over $100$~au. A bright annulus stands out in the inner part of this ring, at around $40$~au from the central star (see Table \ref{TAB:Geometry}).

\section{Methods and results} \label{SECT:Methods}
In this study, we aim to estimate the properties of dust grains in protoplanetary disks by analyzing the SPF. Extracting this measurement requires knowledge of the disk geometry—inclination, position angle, scattering surface height, and flaring—to define the scattering angle, i.e., the angle describing the change of propagation direction between the incident and scattered-light beams (0$\degc$ meaning no change in direction or forward-scattering); and to constrain the analysis within a specific ring of the disk. The observations presented here are scattered-light data, so the flaring refers to the last scattering surface and not to the gas scale height. This study applies the {\tt{DRAGyS}} tool \citep{roumesy_dragys_2025}, which enables a comprehensive and consistent scattered-light analysis, from the geometry to the SPF extraction. This section describes the methods used and the results obtained for the disk geometry and the SPFs and identifies the best dust model for each one.
\begin{figure}
    \centering
    \includegraphics[width=\hsize]{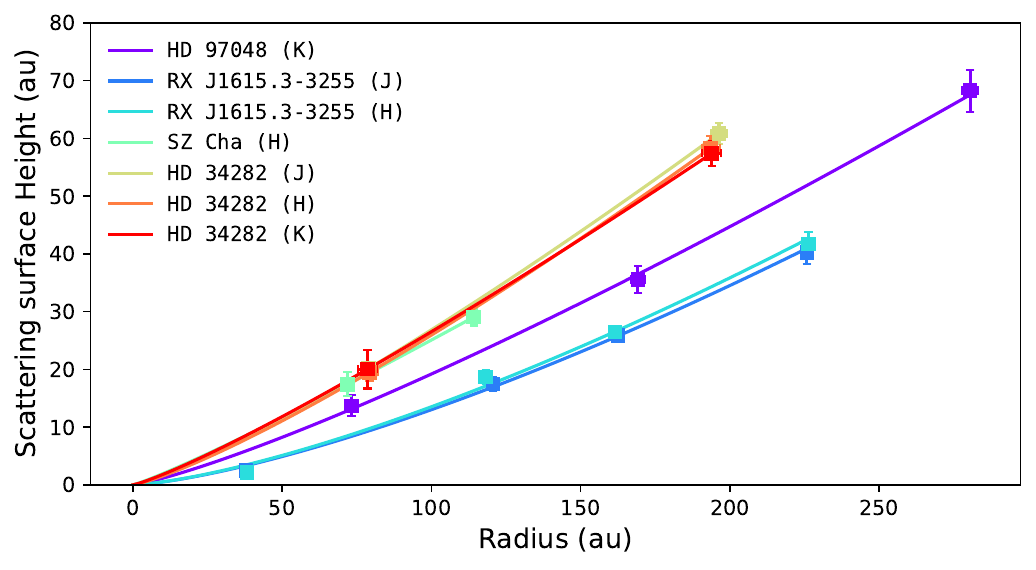}
    \caption{Scattering-surface-height fitting for multiple ring-shaped disks. We plot adjustments as dashed colored curves, and we display each estimated value of scattering height ratio, $h_0/r_0$, and flaring exponent in Table \ref{TAB:Multi_Flaring}.}
    \label{FIG:Multi_Flaring}
\end{figure}

\begin{table}
{\fontsize{8}{10}\selectfont
    \centering
    \caption{Scattering-height estimations for multiple ring-shaped disks.}
    \label{TAB:Multi_Flaring}
    \begin{tabular}{c|c|c|c|c}
    \hline
    Object    &   Band    &   $h_0/r_0$ [au]  &   $r_0$ [au]  &   $\alpha$  \\
    \hline
    HD 97048             & K &   0.217 $\pm$ 0.009   & 174 & 1.23  $\pm$ 0.10  \\[2mm] 
    RX J1615.3-3255      & J &   0.148 $\pm$ 0.005   & 137 & 1.41  $\pm$ 0.11  \\ 
                         & H &   0.153 $\pm$ 0.005   & 136 & 1.41  $\pm$ 0.10  \\[2mm] 
    SZ Cha               & H &   0.249 $\pm$ 0.015   &  93 & 1.11  $\pm$ 0.28  \\[2mm]  
    HD 34282             & J &   0.287 $\pm$ 0.007   & 138 & 1.21  $\pm$ 0.06  \\ 
                         & H &   0.278 $\pm$ 0.010   & 136 & 1.23  $\pm$ 0.09  \\
                         & K &   0.279 $\pm$ 0.019   & 136 & 1.17  $\pm$ 0.19  \\
    \hline
    \end{tabular}}
\end{table}

\subsection{Disk geometry}\label{SECT:Geometry}
Determining the disk geometry is based on the identification of elliptical shapes corresponding to rings within disks to estimate the inclination, position angle, and scattering-height surface, assuming circular axisymmetric rings. This is respectively performed by measuring the minor and major axes of the fitted ellipse, the orientation of the disk, and the offset between the ellipse center and the position of the central star (see \cite{roumesy_dragys_2025} for a detailed description of the {\tt{DRAGyS}} geometric fitting process). We provide an estimation of the geometry for each disk observation, based on this ring detection in Table \ref{TAB:Geometry}.

For disks observed at multiple wavelengths, the surface of last scattering may differ from one band to another, although the observation bands are relatively close to each other. For example, slight deviations of a few degrees are observed in the inclination estimation for the disks around V4046 Sgr, PDS 66, and HD 34282, depending on the wavelength of observation, but they remain within the confidence range. However, there is a significant deviation in the position-angle estimation for UX Tau, which is measured at $339.3\degc \pm 1.8\degc$  in the J band and $355.0\degc \pm 2.8\degc$ in the H band. This offset of more than $15\degc$ is unexpected considering the proximity of the wavelength, but UX~Tau is a specific case as it is affected by a perturber \citep{menard_ongoing_2020}. To ensure consistency between the geometry and the SPFs, we decided to estimate the disk geometry and extract the corresponding SPFs separately for each band.

We estimated the scattering height at a given radius for a ring for each disk, but it is not possible to estimate the disk flaring with only one measurement of height. However, our sample includes several multiple-ring systems, such as HD~97048, HD~34282, RX~J1615.3–3255, and SZ~Cha. We assumed that the scattering surface height as a function of radius is governed by a power law, 
\begin{equation}\label{EQ:Scattering_surface}
    h_s(r) = h_0 \times (r/r_0)^\alpha,
\end{equation}
where $h_0$ is the reference height at a reference radius of $r_0$, and $\alpha$ is the scattering surface flaring exponent. Therefore, we estimated the scattering height surface for each ring of multiple ring systems individually, and we adjusted this $h_s(r)$ function, fixing the reference radius, $r_0$, at the mean radius of each target (see Fig. \ref{FIG:Multi_Flaring}). For the rest of our study, we used the scattering scale height and flaring estimations using this power-law fitting for these multiple-ring systems. 

\begin{figure}
\centering
\includegraphics[width=\hsize]{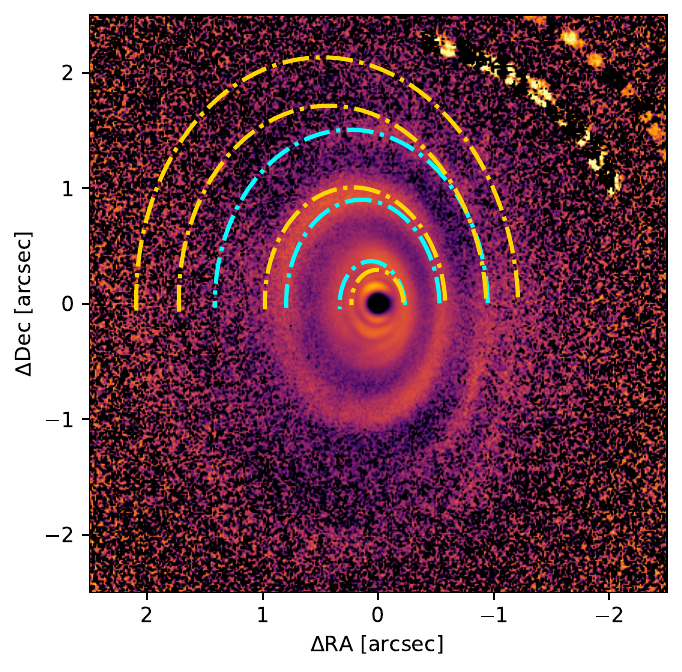}
  \caption{Ring fitting on HD~97048. We compare the estimation presented in \cite{ginski_direct_2016} (in orange) and our estimation (in cyan). The image is on $R^2$-scaled, and we only plot half of the ellipse to bring up substructures. Features to the northwest are artifacts from data reduction.}
     \label{FIG:HD97048_flaring}
\end{figure}

An interesting remark is that the measurement of the scattering height in different bands is very similar, reinforcing the accuracy with which {\tt{DRAGyS}} determines the disk's surface height. This is the case for the disk around HD 34282, with $h_s/r \simeq 0.28$ and $\alpha \simeq 1.22$ in the J and H bands, $\alpha \simeq 1.17$ in the K band, and around RX~J1615.3-3255 with four rings, with $h_s/r \simeq 0.16$ and $\alpha \simeq 1.3$. In contrast, the single-ringed disk around PDS~66 exhibits a decreasing scattering aspect ratio ($h_s/r$) with increasing wavelength, which may indicate that different vertical layers of the disk are being probed. However, given the image noise (particularly in the K-band observations) and the absence of a similar trend in other disks observed across multiple wavelengths, it remains difficult to draw a firm conclusion about the layering of the scattering surface. In the case of the disk around HD~97048, previous studies reported a surprisingly high flaring of $1.73$ \citep{ginski_direct_2016} and suggested that the flux from the disk's back side was mainly dominated by the inner rim of the disk rather than its surface. However, the new observation provides better contrast and reveals a greater part of the disk's back side, leading to a more robust estimation of the disk's geometry, especially to a lower flaring exponent of $1.23$ (see Fig. \ref{FIG:HD97048_flaring}), which is in good agreement with the previous estimation by \cite{lagage_anatomy_2006}.

\begin{figure}
\centering
\includegraphics[width=\hsize]{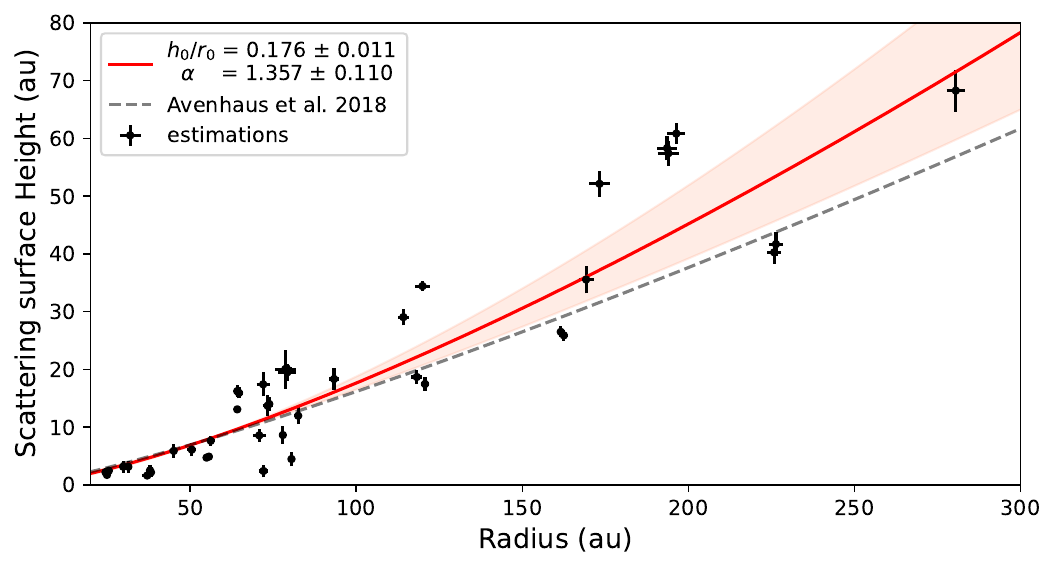}
  \caption{Scattering surface height fitting considering all scattering height estimations. We display fitting in red, and the red zone corresponds to the range of the flaring index at one $\sigma_\alpha$. We display the scattering height estimation from \cite{avenhaus_disks_2018}: $h_s(r) = 16.17~(r/100au)^{1.219}$ measured on a different sample as a dashed gray curve for comparison.}
     \label{FIG:All_Flaring}
\end{figure}

In order to have a global view of the scattering height surface tendencies of protoplanetary disks, we considered all the scattering surface height measurements for all disks simultaneously to adjust the power-law function defined by Eq. \ref{EQ:Scattering_surface}, setting the reference radius at $r_0 = 100$~au. We estimate an average scattering scale height of $h_0/r_0 = 0.176 \pm 0.011$ and an average flaring exponent of $\alpha = 1.36 \pm 0.11$ (see Fig. \ref{FIG:All_Flaring}), consistent with a previous estimation of average scattering scale height ($h_0/r_0 = 0.1617$), but with a stronger disk flaring \citep[$\alpha=1.219$ in][]{avenhaus_disks_2018}. Our sample is larger than in the previous study, resulting in greater dispersion in the surface heights measured for protoplanetary disks. This highlights the wide diversity of disk geometry, which is expected given that the disk's vertical structure depends on many physical parameters (e.g., stellar mass, luminosity, and dust mass) that differ significantly from one system to another. For the rest of our study, we used the individual flaring estimate for disks with multiple rings, and we took this global estimated flaring exponent of $\alpha =1.357$ for the others. 

The overall disk-geometry estimate is sufficiently precise to ensure that the SPF extraction zone is properly positioned on the ring surface. As the scattering surface height estimation was performed directly on the target ring, the influence of flaring on the extraction zone remains minor, especially in the case of narrow rings where scattering height variation is smaller. However, for wider structures, this approximation becomes less reliable, such as in the case of the PDS~70 protoplanetary disk. With a strongly flared ring, measuring the scattering surface height on the inner part of the ring with {\tt{DRAGyS}} is no longer applicable to the entire ring surface. Therefore, to better encompass the ring surface within the extraction zone in the case of PDS~70, we manually increase the flaring value to $1.6$ and the scattering scale height to $0.175$ rather than the estimated value of $0.135$. We can note the same feature for TYC~6213-1122-1. It also has a relatively wide ring, but with a very bright region in the inner part. For this reason, we decided to focus the analysis on this bright region rather than the entire ring. 

As an additional point, the disk flaring estimation based on the estimated scattering height for each ring does not necessarily follow the theoretical flaring expected for a smooth, continuous disk. Indeed, this flaring is well established for protoplanetary disks without substructures \citep{chiang_spectral_1997, dullemond_passive_2001}, but the formation of rings and gaps can significantly impact the vertical structure of the disk, particularly through variations of dust mass within each ring, which affects the scattering height from ring to ring. Consequently, the optical-depth surface, $\tau=1$, observed on each ring does not necessarily reproduce the global flaring of the disk. All of our flaring estimations are greater than 1, which is consistent with the geometric requirements for the observation of such rings. Curiously, the flaring values derived from the scattering surface height of each ring, as in our study, are consistent with the case of a continuous disk-flaring exponent, suggesting an underlying similarity between the surface properties of disks with and without marked substructures.

\subsection{Phase-function extraction} \label{SECT:SPF}
\begin{figure*}[ht]
    \centering
    \includegraphics[width=\textwidth]{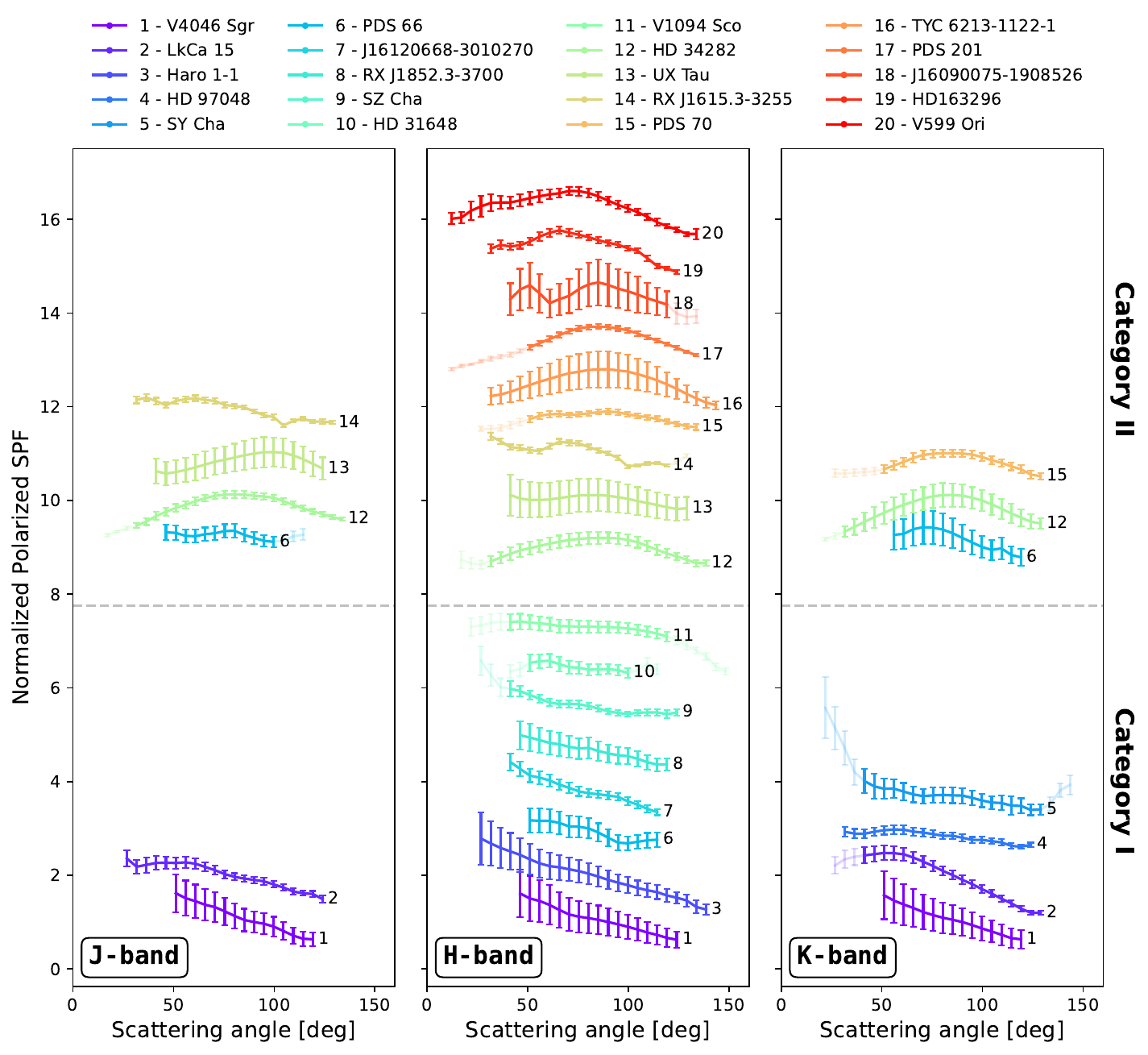}
    \caption{Extracted SPFs from our datasets for J, H, and K bands. All SPFs are shifted vertically for clarity. The colors correspond to observed systems.}
    \label{FIG:SPFs}
\end{figure*}
The next step is the extraction of the SPFs, which was also performed with the {\tt{DRAGyS}} tool \citep{roumesy_dragys_2025}. We focused our study on a specific area of the disk's surface, usually a ring, by defining an extraction zone delimited by two concentric circles with radii of $R_{in}$ and $R_{out}$; these were projected according to the previously estimated disk geometry (see Table \ref{TAB:Geometry}). All SPFs were normalized to a $90\degc$ scattering angle, and the uncertainties include the errors on the estimated geometry, combined with the standard deviation within each bin. For the rest of our study, especially for adjusting the SPFs using the dust models presented later in Sect. \ref{SECT:Dust}, we decided to add a floor uncertainty of $0.05$ to all SPF points as an estimate of the systematic uncertainty of the method.

For multiple-ring systems, we present the SPFs extracted for each ring in Appendix \ref{APP:allSPF}. The SPF shape varies from one ring to another, which could reflect a radial grain-size distribution. However, this point is beyond the scope of the paper and will be explored in a future study. We focused the study on the best-defined ring for the following statistical study. In the case where the target is observed at several wavelengths, we keep the same extraction zone limits $R_{in}$ and $R_{out}$ for all wavelengths. The extraction zone covers the entire azimuth of the ring. However, it could overlap the coronagraph in some cases, or the transmission of the coronagraph could contaminate the surface brightness due to large inclination or strong flaring of the disk. This is the case for HD~34282, PDS~201, and V~599~Ori, where the SPFs are particularly affected for small scattering angles. To avoid such contamination, we decided to exclude the lower scattering angle for these SPFs. In some cases, other parts of the disk needed to be excluded to preserve the SPF. As an example, PDS 66 observed in the K band shows that the farthest part of the disk appears strongly affected by background noise. It may be useful to exclude the extracted SPF from these scattering angles. However, to keep the SPFs as consistent as possible, we decided to ignore this last point. This should be kept in mind when analyzing SPFs.

As a last step, the protoplanetary disk's surface brightness is affected by the limb brightening effect, making the forward side of the disk brighter than the back side \citep{tazaki_fractal_2023, ginski_observed_2023, roumesy_dragys_2025}. This purely geometric effect alters our extracted SPFs and must be corrected for in our measurements to better evaluate the intrinsic dust SPF lying on the disk's surface. All corrected SPFs are displayed in Fig. \ref{FIG:SPFs}, although only the one from the well-defined ring is shown for the multiple ring-shaped systems. More detailed SPFs are presented in Appendix \ref{APP:allSPF}. 
A clear distinction is identified between two shapes of SPFs in \cite{ginski_observed_2023}. They define Category I as presenting a monotonic decrease in SPF with the scattering angle, whereas a bump characterizes Category II. However, when considering a larger sample of disks and after correcting for limb brightening that tends to flatten the SPFs, it appears that the observed SPF shapes follow a more continuous distribution. For simplicity (and direct comparison), and because we aim to identify (broad) families of dust shapes in protoplanetary disks, we retained this division into two categories of SPFs at this point. As a consequence, in some cases, this classification remains ambiguous. For example, PDS~66 shows a small bump in the J band, so we placed it in Category II, although it was classified as Category I by \cite{ginski_observed_2023}. Conversely, RX~J1615.3-3255 exhibits an almost monotonic shape in the J band, but a marked bump in the H band. So, its first Category II classification is maintained \citep{ginski_observed_2023}. The disks around PDS 70 and UX Tau present relatively flat profiles in the H band, but reveal a bump in the J band for UX Tau and in the K band for PDS 70. We therefore classified them as Category II. Lastly, targets such as HD 97048, HD 31648, and V1094 Sco also have relatively flat shapes, but their detailed SPFs (see Appendix \ref{APP:allSPF}) indicate a monotonic decrease, justifying their inclusion in Category I.

\subsection{Dust properties} \label{SECT:Dust}
To determine which dust model better explains the SPFs from category I and category II, we used the {\tt{AggScatVIR}} (Aggregate Scattering for Visible and Infrared wavelengths) database, developed by \citep{tazaki_size_2022, tazaki_fractal_2023}, where they computed the theoretical SPFs for 360 different dust models. The database is divided into three types of particles. The first type is compact aggregates formed by at least eight and at most 4096 equally sized monomers ($100$ nm, $200$ nm, or $400$ nm), providing a maximum grain size ranging from $0.22$ to $3.14~\mu$m and an equivalent volume radius between $0.2$ and $1.6~\mu$m. Compact aggregates are divided into three categories: low porosity (CALP), medium porosity (CAMP), and high porosity (CAHP). The second type is fractal grains, with a fractal dimension of $D_f = 1.1,~1.3,~1.5$, and $1.9$, composed similarly to compact aggregates, but with additional monomers of $150$~nm and $300$~nm in size. The fractal grains offer a wider range of maximum grain size (from $3.6$ to $12~\mu$m), and the corresponding equivalent volume radius ranges from $0.2$ to $2.0~\mu$m. Finally, irregular grains consisting of a Gaussian random sphere (GRS) are defined by their equivalent radius, which ranges from $0.2$ to $1.6~\mu$m. 

{\tt{AggScatVIR}} allows optical properties to be averaged over a particle size distribution $n(a_v)da_v=a_v^{-3.5}da_v$, where $a_v$ is the volume-equivalent radius of particles. Hence, the maximum grain size of particles is larger than the size specified by this value for higher porosity,  \citep{tazaki_fractal_2023}. This distribution is a discrete sampling based on the number of monomers but not their size and on the volume-equivalent radius for the Gauss. Particle size distribution is applied throughout the rest of the study to better approximate the dust population in protoplanetary disks, and we also performed the same analysis when turning off this size distribution in Appendix \ref{APP:Statistics_single}. {\tt{AggScatVIR}} provides two different compositions. Both are a mixture of water ice (20\% in mass), silicate (32.9\%), and troilite (7.4\%), to which either amorphous carbon \citep[amc,][]{zubko_optical_1996} or organic matter \citep[org,][]{henning_dust_1996} is added.

\begin{figure}
\centering
\includegraphics[width=\hsize]{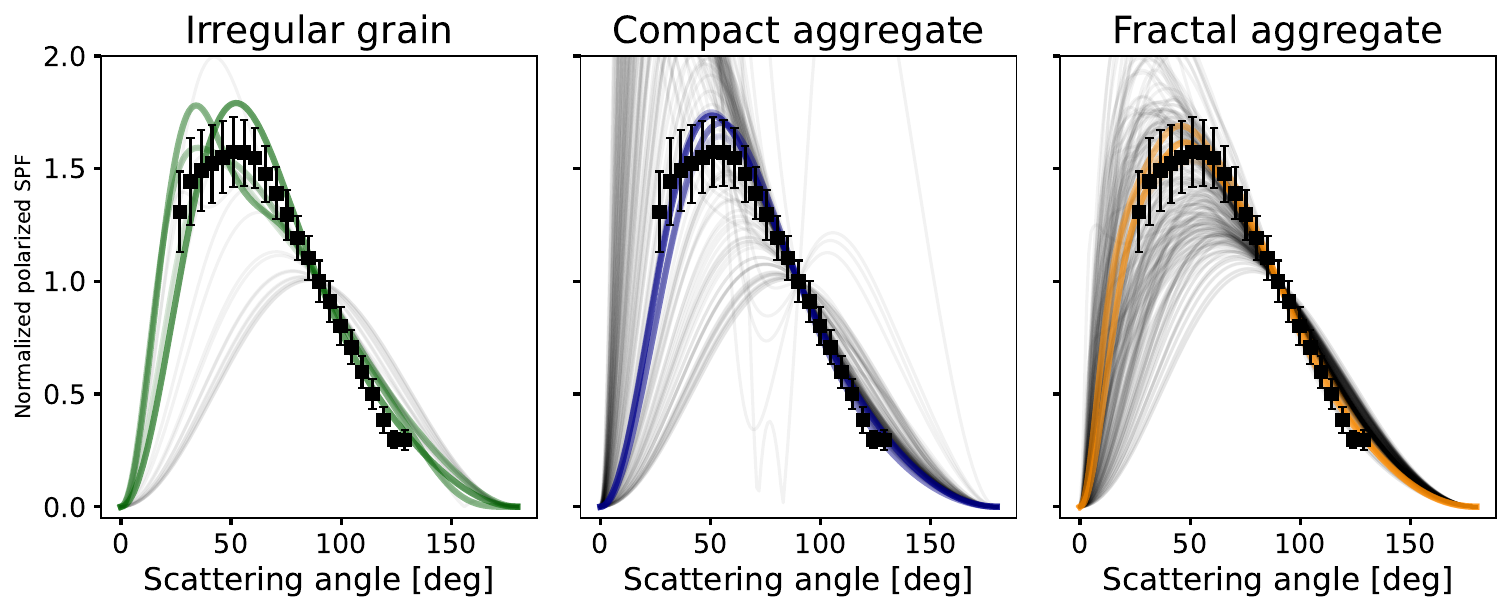}
    \caption{SPF fitting for LkCa~15 in the K band using the three families of dust models from {\tt{AggScatVIR}}: irregular grains constructed by a GRS and compact and fractal aggregates (from left to right). The five best models are shown as colors, and all others are displayed as faint dark curves.}
    \label{FIG:LkCa15_Example}
\end{figure}

\begin{figure*}[ht]
    \centering
    \includegraphics[width=\textwidth]{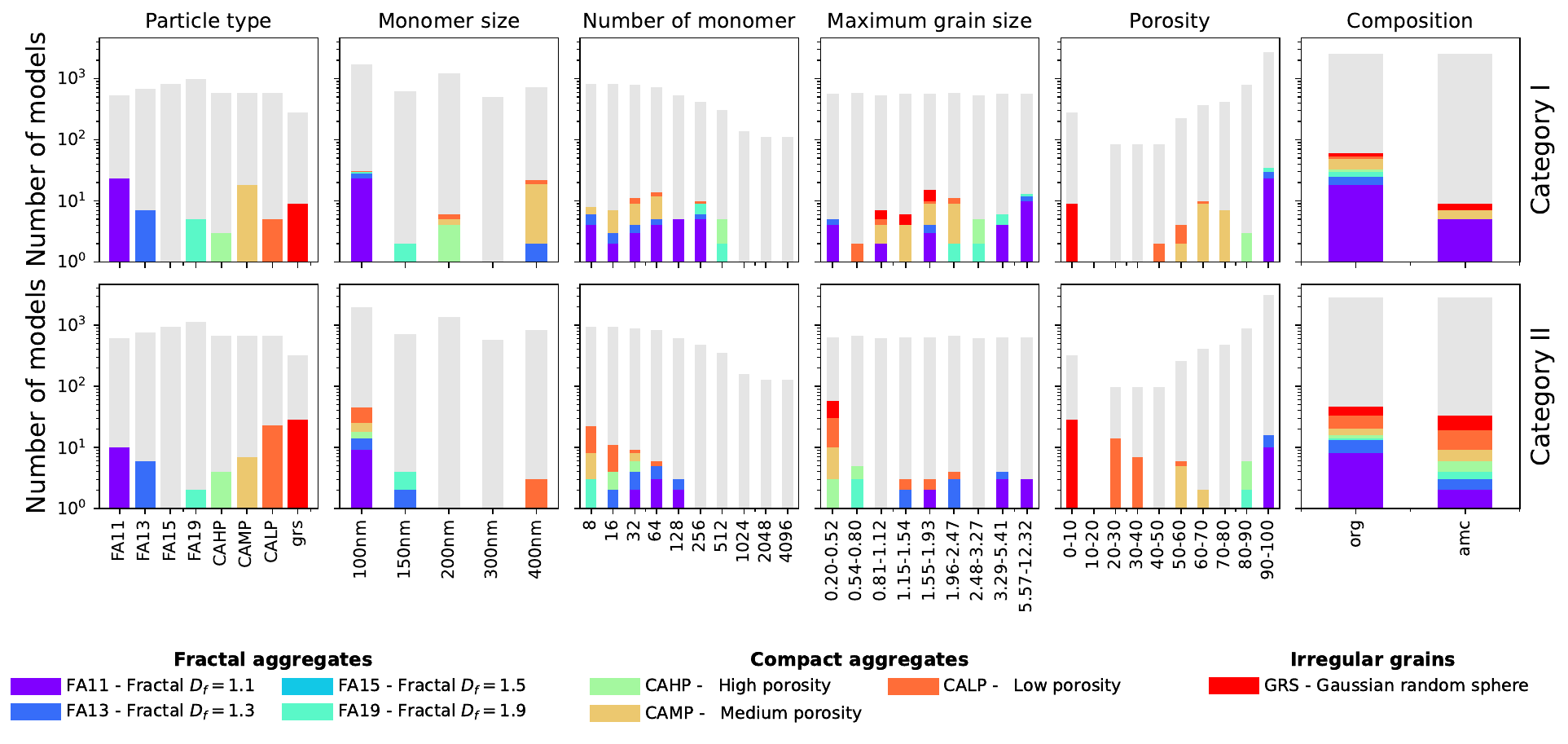}
    \caption{
    Statistical panel of best models estimated from the Category I (top) and Category II (bottom) SPFs. The colors are linked to the type of particles from {\tt{AggScatVIR}} and displayed in the left column. We kept the same colors for the other histogram to track their distribution in monomer size, number of monomers, maximum grain size, porosity, and composition (from left to right). The gray bars represent the total number of all models considered to show the range of possibilities. Note that irregular grains (in red) are not composed of monomers, so they are not visible in the histogram related to aggregates. Furthermore, as they are dense particles, we fixed their porosity to 0.}
    \label{FIG:Statistics}
\end{figure*}

For each category, the SPFs extracted from the disk presented in Sect. \ref{SECT:SPF} are compared to the theoretical SPFs from the {\tt{AggScatVIR}} dust models. The compatibility between observations and models is assessed using a $\chi^2$ adjustment, and only the five models with the lowest values are selected for each case. This fitting process is shown for the disk around LkCa~15 in Fig. \ref{FIG:LkCa15_Example}. This approach aims to identify families of dust models to perform a more representative statistical analysis. Thus, for each category, the retained dust models are combined, and the global distribution of results is presented in the histograms in Fig. \ref{FIG:Statistics}.

The statistics obtained on the SPFs in Category I reveal that fractal aggregates with $D_f=1.1$ and very small monomers of $100$~nm are favored, with a strong preference for organic matter composition. CAMPs are also a reasonable candidate, with large monomers of $400$~nm offering a maximum grain size comparable to the observational wavelengths. The two candidates also differ in their porosity, as fractals have a high porosity greater than 90\% while CAMPs are more compact, with a porosity around 70\%. For category II, SPFs fitting shows a predominance of irregular grains or small compact aggregates, both consistent with a maximum grain size smaller than $0.5\mu$m. However, organic and amorphous carbon compositions are almost equally probable. The study of SPF shapes in about thirty observations of protoplanetary disks extends the first analysis proposed in \cite{ginski_observed_2023}. Although a continuous distribution of SPF shapes seems to emerge, it appears that the distinction between the two categories provides a fairly clear identification of two families of dust grains in protoplanetary disks: porous aggregates and small compact grains.

\section{Discussion} \label{SECT:Discussion}
\subsection{SPF shapes: Category I}
To describe the first category of SPF, the best dust models clearly show a predominance of aggregate-shaped grains, especially with medium porosity (CAMP). To describe the first category of SPF, the best dust models clearly show a predominance of fractals. The best model is defined by a fractal dimension of $D_f = 1.1$, composed of organic matter with a monomer size of $100$~nm, and very high porosity (greater than 90\%). Furthermore, no general trend emerges for the maximum grain size (i.e., number of monomers), suggesting that the SPF provides more constraints on the monomer size within the aggregate than on the overall size of the dust particle. This global analysis to explain category I SPFs generalizes the results obtained by \cite{tazaki_fractal_2023}, whose study of the SPF in IM Lup also suggests a fractal dust model. However, this study concluded that a highly absorbing composition (amorphous carbon) is needed to explain the SPF of IM Lup. Because the study of \cite{tazaki_fractal_2023} is for an individual case, it is perhaps not surprising that some differences arise compared to our more global trend.

In addition to fractals, compact aggregates may also be reasonable candidates to explain the shape of SPFs from Category I. Especially compact aggregate with medium porosity, made of large monomers measuring approximately $400$~nm, with a total number ranging from 8 to 64 monomers. This results in a preferred maximum grain size ranging from $0.2~\mu$m to nearly $3~\mu$m. This wide range of grain sizes suggests that the SPF of aggregates depends more on the size of the monomers than on that of the overall particle. In other words, the shape of the particle, defined by its fractal dimension or porosity, strongly influences the dependence of the scattered light on the specific properties of the dust grains. Specifically, the SPF of a spherical dust particle or an irregular grain depends mainly on the overall size of the particle. In contrast, for compact porous aggregates or more complex fractal structures, the SPF is more influenced by the size of the monomers that compose the aggregate than by its overall size. A more thorough exploration of the influence of particle shape on SPF is therefore required to better understand these dependencies. 

\subsection{SPF shapes: Category II}
For the second category of bell-shaped SPFs, the most promising dust models are irregular grains (GRS) and compact aggregates with low porosity (CALP), both with a maximum grain size of less than $0.5~\mu$m. These two models do not exhibit a strong preference for any composition. Although there seems to be a slight tendency toward amorphous carbon (amc), the shape and porosity of the grains seem to govern the SPF more than their composition. Overall, sub-micron-sized dust particles that are highly compact generally explain SPFs with a bump.

\begin{figure}
    \centering
    \includegraphics[width=0.99\linewidth]{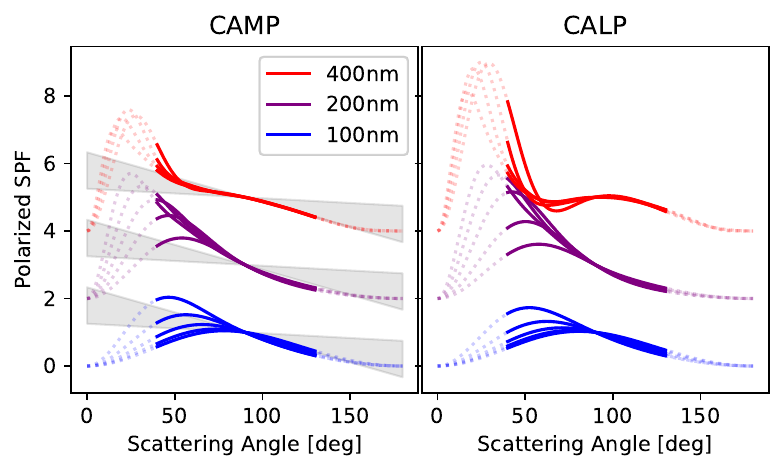}
    \caption{Theoretical SPFs of compact aggregates with medium (left) and low (right) porosity for different monomer sizes, shifted for clarity. Solid lines represent the typical range of scattering angles available in protoplanetary-disk observations, while dashed lines indicate the inaccessible range. For aggregates with medium porosity (a dust model that may explain Category I scattering angles), a gray area indicates the range of SPFs for Category I.}
    \label{FIG:Monomers}
\end{figure}

An additional finding is the bimodal distribution of monomer sizes observed in CALP-type dust models, for which no solution was found at $200$~nm in our sample. This is due to the lack of a bell-shaped profile in the SPF of the CALP models at $200$~nm, unlike at $100$~nm and $400$~nm, where a bump is clearly visible (see Fig. \ref{FIG:Monomers}). This behavior goes beyond the distinction of SPF in categories; a similar trend is observed in Category I, where the two preferred dust models are clearly split between $100$~nm for FA11 and $400$~nm for CAMP due to the characteristic slope of their respective SPFs. The distinction observed in monomer sizes is the result of an observational bias. Current instruments and the disk geometry do not provide access to lower scattering angles. As shown in Fig. \ref{FIG:Monomers}, this part of the SPF could exhibit a bump, as is the case for $200$~nm CALP, and access to these smaller angles would provide a much more robust constraint on the dust properties derived from SPFs.

\subsection{SPF asymmetry}
Some SPFs from Category II show strong asymmetry coming from the disk's surface brightness. A typical example in our sample is HD~163296, in which a significant brightness enhancement is visible on the north-west side. If we separate the contribution of SPFs from the east and west sides along the disk's minor axis, we clearly identify that one side has a strong bump at around $70\degc$ and belongs to Category II, whereas the other one has a decreasing shape and belongs to Category I. In our study, we assume that the surface brightness is dominated by scattering by dust grains well mixed in the disk. However, this asymmetry in SPFs, also visible in the disks around HD~97048, HD~34282, PDS~70, PDS~201, RX~J1615.3-3255, and V599 Ori, breaks our assumption for these disks, as the surface brightness is affected by physical processes beyond the properties of dust and raises the question of which disk portion can be used to estimate the intrinsic dust SPFs. These asymmetries may arise from the presence of planetary companions within the disks, which induce geometric perturbations such as warps, shadowing, or misalignment of the inner disk. This alters the distribution of scattered light on the disk surface, leading to asymmetric SPFs unsuitable for estimating the intrinsic properties of dust grains. Within our sample, several disks have been studied to detect or predict such companions, such as PDS~70 \citep{keppler_discovery_2018, blakely_james_2025}, RX~J1615.3-3255 \citep{de_boer_multiple_2016, avenhaus_disks_2018, asensio-torres_perturbers_2021}, HD~97048 \cite{pinte_kinematic_2019}, and HD~163296 \citep{pinte_kinematic_2018, izquierdo_circumplanetary_2026}. To estimate the properties of dust in such cases, a radiative-transfer model is required to disentangle the intrinsic dust SPF from other contributions. We repeated the statistical analysis described in Sect. \ref{SECT:Dust} to identify the best dust model for each category, excluding SPFs with strong asymmetry. This selection reduced our Category II sample but does not change the main results obtained before. It even reinforces the dominance of very compact sub-micrometric grain models such as GRS or CALPs, explaining the shape of Category II SPFs, as fractal grain models ($D_f = 1.1$ and $1.3$) present in the initial statistics (see Fig. \ref{FIG:Statistics}) disappear once the asymmetric SPFs are excluded.

\subsection{Individual source: Degeneracy}
This study highlights general trends regarding the best dust model to reproduce the shape of SPFs in each category, but individual interpretation may remain highly degenerate for some systems. This is the case for the disk around LkCa~15, for example (see Fig. \ref{FIG:LkCa15_Example}); while fractal aggregates are the favored candidates, some compact aggregate models can also provide a credible fit. Nevertheless, the shape of the SPF still imposes useful constraints on some properties within each category, such as trends in maximum grain size, porosity, or composition. In future studies, it will be necessary to break down this degeneracy to examine the dust within the disks individually. One effective strategy is to use multiwavelength observations to extract SPFs and measure the color of the disk, which is directly related to the optical properties of dust grains and independent of the disk geometry. This is the approach adopted for IM~Lup in \cite{tazaki_fractal_2023}, where combining several bands allows us to link the shape of the SPFs and the disk color to a more restricted set of grain properties. Moreover, the wavelength dependence of the SPF and color is inherent to the dust model. So, comparing the same model to multi-band SPFs simultaneously adds constraints consistent across wavelengths and can break the degeneracy encountered when using a single SPF.

\subsection{Dust growth and origin of aggregates}
The growth of dust in protoplanetary disks begins with so-called first-generation sub-micronic grains originating from the interstellar medium, whose low relative collision velocities in the protoplanetary disk promote “hit-and-stick” coagulation, forming fractal aggregates with $D_f\leq2$ \citep{blum_growth_2008}. As the aggregates become larger, collision velocity increases, leading to hit-and-stick collisions with deformation, while maintaining high porosity \citep{blum_growth_2008, testi_dust_2014}. When velocities increase further, compaction, fragmentation, or erosion phenomena occur, still producing porous aggregates but not fractals \citep{blum_growth_2008, birnstiel_dust_2016}. The most suitable dust models for SPF categories suggest that Category II (compact sub-micronic grains) corresponds to the initial growth phase with small compact grains from the ISM, while Category I (porous fractal or compact aggregates) may correspond to a more advanced growth stage involving compaction and/or fragmentation \citep{blum_growth_2008, testi_dust_2014}. However, since our sample consists of disks with a ring structure, often interpreted as more advanced systems where planet formation might be occurring, it is plausible that the two categories of SPFs, thus the two associated dust families, reflect distinct stages of the planet formation process. \citep{turrini_dust--gas_2019}. \cite{bernabo_dust_2022} shows that the amount of dust within the disks either remains stable or increases during the first 1 to 3 Myr. This is consistent with a scenario of collisions between planetesimals excited by the planet formation process that increases the dust population and generates a second generation of sub-$\mu$m-sized dust particles \citep{bernabo_dust_2022, birnstiel_dust_2024}. Given current state-of-the-art methods, we cannot distinguish between the first generation of dust originating from the ISM---suggesting a very young disk without a planetary core---and second-generation dust, which suggests the presence of planetesimals or even a planet, depending on the rate of planetesimal formation \citep{birnstiel_dust_2016, bernabo_dust_2022, tatsuuma_bulk_2024}. This explains the lack of correlation between the age of the system and the SPF category in our sample, as already mentioned in \cite{ginski_observed_2023}, despite the influence of turbulence on collision velocities, which affects dust growth \citep{birnstiel_dust_2024}.

\begin{figure}
    \centering
    \includegraphics[width=0.99\linewidth]{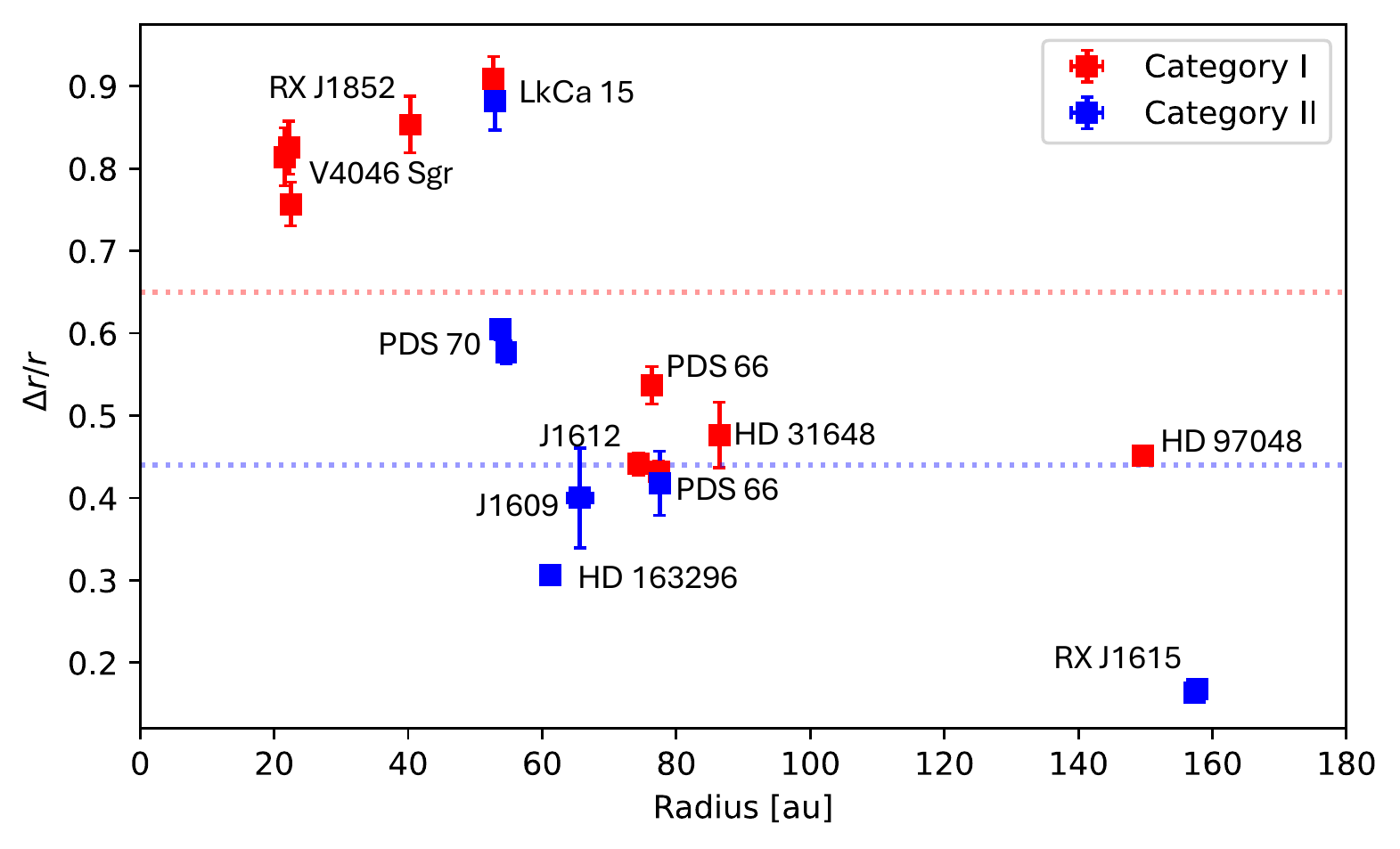}
    \caption{Ratio between ring width and radial distance from the central star for 18 out of 30 disk observations. Measurements are separated according to the SPF-based classification: red markers indicate Category I, and blue markers indicate Category II. The dashed colored lines represent the mean value of $\Delta r/r$ for each category, where $\Delta r$ is the FWHM of the fitted Gaussian.}
    \label{FIG:RingWidth_vs_Category}
\end{figure}

As an additional step, we tried to find a connection between the geometry and the category of dust we found using SPFs. As expected, we do not observe any connection between the dust families and the inclination of the disk, as well as for the position angle and the scattering scale height of the disk. We also tried it with the stellar spectral type, and again, no correlation appeared. The last test we carried out was on the ring structure of the disk. We estimated the ring width and its radius to the star by fitting the disk's radial profile with a Gaussian for some disks in our sample where the ring structure is well defined (see Fig. \ref{FIG:RingWidth_vs_Category}). As a general view, the ratio between ring width and the radius seems to be smaller for the small compact grains (i.e., Category II) than for the aggregate particles (i.e., Category I). We applied the Mann--Whitney and Kolmogorov--Smirnov statistical tests to confirm the differences in ring widths, with confidence levels of 94\% and 97\%, respectively.

Assuming that rings in protoplanetary disks indicate a relatively advanced stage of planetary formation, this remark suggests that the narrow rings could host a population of second-generation small compact grains formed after an initial phase of planet or planetesimal formation \citep{tatsuuma_bulk_2024}. Conversely, broader rings would contain more porous grains such as aggregates, as they may still be in a dust-growth phase induced by coagulation. However, the observed trend does not constitute a strict correlation, and it remains difficult to draw definitive conclusions about the ring geometry and the dust population.

\section{Summary}

In this paper, we present a statistical analysis of a large sample of thirty $Q_{phi}$ polarized-intensity images of protoplanetary disks, observed in the near infrared by the VLT/SPHERE instrument. We used the {\tt{DRAGyS}} tool to measure the geometry of all these ring-shaped disks and extract the polarized SPFs corrected for the limb-brightening effect. The key results are listed below.
\begin{itemize}
    \item The SPF categorization, defined in \citep{ginski_observed_2023}, was validated and completed according to the SPF shape. Category I shows decreasing SPFs, while Category II is distinguished by a bell shape of around $60-80\degc$ .
    \item Combining all scattering surface height measurements provides a global estimation of the disk flaring, with a power-law exponent of $\alpha=1.357~\pm~0.15$, consistent with the value reported in a similar previous study \citep{avenhaus_disks_2018}.
    \item The disks around HD~97048, RX~J1615.3-3255, SZ~Cha, and HD~34282 have several rings, allowing for the individual estimation of their flaring, ranging from $1.08$ to $1.38$. For disks observed at multiple wavelengths, the scattering surface height does not significantly change from one band to another.
\end{itemize}

This study proceeded with an estimation of the best dust models for each SPF category, using the {\tt{AggScatVIR}} database \citep{tazaki_size_2022, tazaki_fractal_2023}, containing a variety of dust-particle shapes. A $\chi^2$ adjustment between these models and the extracted SPFs enabled us to identify some general trends:

\begin{itemize}
    \item Category I SPFs are dominated by fractal dust models composed of small monomers of $100$~nm, with very high porosity (greater than 90\%) and a composition clearly dominated by organic matter (org). However, compact aggregates (CAMP) seem to be a reasonable candidate too, with large monomers ($400$~nm) and lower porosity (70\%). Although the size of compact aggregates appears to be similar to the observation wavelength (from $\sim1$ to $3~\mu$m), Category I SPFs are preferentially affected by monomer size, composition, and particle shape rather than grain size.
    \item Category II SPFs are clearly dominated by small sub-micrometric dust particles with irregular shapes or low-porosity dust aggregates. This category shows significant dependence on particle size and porosity, but not so much on dust composition. 
    \item Furthermore, strong asymmetry is often observed in Category II SPFs. For example, the disk around HD~163296 shows SPFs of different categories on each side of the disk. These asymmetries could result from various mechanisms such as dust traps in pressure bumps, vortices, and alignment issues between the inner and outer disks, or even interactions with planetary companions forming within the disk. However, current observations are not sufficient to conclude which mechanism is correct or to infer the presence of planets based on these asymmetric SPFs.
    \item Dust models for each category can probe different stages of the grain growth scenario, as porous aggregates are formed after a hit-and-stick collision between small particles and can explain the Category II SPFs. However, two scenarios can explain the small compact grains for Category I SPFs. The first one is that sub-micron grains may originate from the ISM and be directly incorporated during the early stage of grain growth. Alternatively, the collision between planetesimals could fragment them, releasing these dense sub-micron grains \citep{tatsuuma_bulk_2024}. The ring-shaped structure of the disk may be caused by various mechanisms. One of these involves an advanced stage of planetary formation and could lend more credibility to the second scenario.
\end{itemize}

We emphasize that this statistical study provides a general trend on the dust properties in disks and will guide future studies on dust populations existing within disks. However, a similar study on a single disk is not enough to accurately evaluate the properties of the dust grains it contains, due to the high degeneracy in the results. To overcome this degeneracy, a multiwavelength study is required, not only to obtain multiple SPFs but also to access information about the disk's color, which is directly related to the optical properties of the dust and independent of the disk geometry. Among our sample, some polarized intensity color was already estimated from NIR observations of disks such as V4046~Sgr, PDS~70 \citep[despite its notable temporal variability][]{ma_temporal_2024}, PDS~66, LkCa~15, and RXJ 1615.3-3255 \citep{ma_color_2024, avenhaus_disks_2018}. 

The key to fully exploiting these additional constraints will be to understand precisely how the disk's color varies as a function of different dust models. This is particularly important for the aggregate models simulated with {\tt{AggScatVIR}} \citep{tazaki_fractal_2023, tazaki_size_2022}, which remain the most plausible candidates for populating protoplanetary disks. 

\begin{acknowledgements}
This project has received funding from the European Research Council (ERC) under the European Union’s Horizon Europe research and innovation program (grant agreement No. 101053020, project Dust2Planets, PI: F. M\'enard). RT was supported by JSPS KAKENHI grant Number JP25K07351.
\end{acknowledgements}

\bibliography{Biblio}
\appendix
\onecolumn

\newpage
\section{Geometric parameters of the disks}
In this section we present the geometric parameters estimated using {\tt{DRAGyS}} \citep{roumesy_dragys_2025} for all of the protoplanetary disk observations presented here. These results are listed in Table \ref{TAB:Geometry}.

\begin{table*}[htbp]
\caption{Geometric parameters estimated to extract the SPF from the surface of protoplanetary disks.}
\label{TAB:Geometry}
\parbox{\textwidth}{
\centerline{\begin{tabular}{@{\extracolsep{2.0mm}} c c c c c c c c}
\toprule\midrule
Object & Distance [pc] & Band & PA [deg] & i [deg] & $h/r$ & flaring & $R_{ex}$ [au]  \\
\midrule
V4046 Sgr            & 71.5    & J &    82.6 $\pm$ 4.2     &    35.0 $\pm$ 2.4     &   0.097 $\pm$ 0.022   & 1.357   &    23 - 34    \\
                     &         & H &    73.9 $\pm$ 4.2     &    38.1 $\pm$ 1.9     &   0.069 $\pm$ 0.020   & 1.357   &    23 - 34    \\
                     &         & K &    74.8 $\pm$ 6.0     &    34.3 $\pm$ 3.3     &   0.090 $\pm$ 0.027   & 1.357   &    23 - 34    \\[1mm] 
PDS 66               & 97.9    & J &   187.8 $\pm$ 3.2     &    35.0 $\pm$ 1.8     &   0.145 $\pm$ 0.018   & 1.357   &    70 - 95    \\
                     &         & H &   189.2 $\pm$ 3.8     &    31.9 $\pm$ 2.0     &   0.111 $\pm$ 0.020   & 1.357   &    70 - 95    \\
                     &         & K &   187.2 $\pm$ 3.5     &    31.9 $\pm$ 1.7     &   0.056 $\pm$ 0.016   & 1.357   &    70 - 95    \\[1mm] 
LkCa 15              & 157.2   & J &    61.6 $\pm$ 2.1     &    50.1 $\pm$ 1.3     &   0.253 $\pm$ 0.016   & 1.357   &    60 - 80    \\
                     &         & K &    58.2 $\pm$ 2.6     &    51.2 $\pm$ 1.6     &   0.246 $\pm$ 0.016   & 1.357   &    60 - 80    \\[1mm]
PDS 70               & 112.4   & H &   341.7 $\pm$ 1.7     &    52.7 $\pm$ 0.9     &   0.175 $\pm$ 0.009   & 1.600   &    50 - 90    \\
                     &         & K &   338.6 $\pm$ 2.8     &    50.8 $\pm$ 1.7     &   0.175 $\pm$ 0.016   & 1.600   &    50 - 90    \\[1mm] 
UX Tau               & 142.2   & J &   349.9 $\pm$ 5.5     &    41.4 $\pm$ 3.2     &   0.101 $\pm$ 0.032   & 1.357   &    30 - 45    \\
                     &         & H &   352.9 $\pm$ 6.4     &    44.9 $\pm$ 3.9     &   0.106 $\pm$ 0.036   & 1.357   &    30 - 45    \\[1mm] 
HD 34282             & 308.6   & J &   295.9 $\pm$ 2.2     &    59.2 $\pm$ 1.2     &   0.256 $\pm$ 0.011   & 1.212   &    65 - 100   \\
                     &         & H &   294.9 $\pm$ 4.0     &    60.8 $\pm$ 2.0     &   0.246 $\pm$ 0.020   & 1.227   &    65 - 100   \\
                     &         & K &   293.9 $\pm$ 6.0     &    55.0 $\pm$ 4.5     &   0.255 $\pm$ 0.044   & 1.166   &    65 - 100   \\[1mm] 
RX J1615.3-3255      & 155.6   & J &   146.9 $\pm$ 0.9     &    48.7 $\pm$ 0.6     &   0.159 $\pm$ 0.006   & 1.378   &   145 - 180   \\
                     &         & H &   146.4 $\pm$ 0.9     &    48.4 $\pm$ 0.6     &   0.164 $\pm$ 0.006   & 1.340   &   145 - 180   \\[1mm] 
HD 97048             & 184.4   & K &     0.2 $\pm$ 2.1     &    45.3 $\pm$ 1.3     &   0.204 $\pm$ 0.014   & 1.23   &   155 - 195   \\[1mm] 
Haro 1-1             & 151.1   & H &   225.0 $\pm$ 3.8     &    54.0 $\pm$ 2.4     &   0.121 $\pm$ 0.023   & 1.357   &    45 - 70    \\[1mm] 
J16120668-3010270    & 132.0   & H &   218.0 $\pm$ 2.0     &    37.6 $\pm$ 1.5     &   0.190 $\pm$ 0.016   & 1.357   &    70 - 90    \\[1mm] 
RX J1852.3-3700      & 147.1   & H &   292.1 $\pm$ 3.4     &    36.5 $\pm$ 2.5     &   0.131 $\pm$ 0.025   & 1.357   &    40 - 60    \\[1mm] 
SZ Cha               & 190.2   & H &   335.9 $\pm$ 1.8     &    47.5 $\pm$ 1.2     &   0.254 $\pm$ 0.013   & 1.108   &   100 - 140   \\[1mm] 
HD 31648             & 156.3   & H &   325.6 $\pm$ 2.5     &    37.8 $\pm$ 1.9     &   0.197 $\pm$ 0.020   & 1.357   &    80 - 110   \\[1mm] 
V1094 Sco            & 154.8   & H &   104.2 $\pm$ 2.7     &    61.4 $\pm$ 0.9     &   0.088 $\pm$ 0.010   & 1.357   &    50 - 85    \\[1mm] 
SY Cha               & 180.7   & K &   350.4 $\pm$ 3.7     &    59.0 $\pm$ 1.5     &   0.121 $\pm$ 0.016   & 1.357   &    60 - 90    \\[1mm] 
TYC 6213-1122-1      & 154.6   & H &    31.8 $\pm$ 4.1     &    57.3 $\pm$ 2.2     &   0.044 $\pm$ 0.017   & 1.357   &    30 - 50    \\[1mm] 
PDS 201              & 323.8   & H &   136.4 $\pm$ 2.0     &    62.3 $\pm$ 0.8     &   0.287 $\pm$ 0.009   & 1.357   &   100 - 140   \\[1mm] 
J16090075-1908526    & 137.4   & H &   330.4 $\pm$ 2.6     &    47.9 $\pm$ 1.7     &   0.034 $\pm$ 0.015   & 1.357   &    55 - 75    \\[1mm] 
HD163296             & 101.0   & H &   134.0 $\pm$ 1.4     &    46.8 $\pm$ 1.0     &   0.204 $\pm$ 0.010   & 1.357   &    55 - 75    \\[1mm] 
V599 Ori             & 401.1   & H &   313.4 $\pm$ 2.3     &    58.8 $\pm$ 1.2     &   0.301 $\pm$ 0.014   & 1.357   &   150 - 200   \\[1mm] 
\midrule
\end{tabular}}}
\tablefoot{The position angle (PA) is defined as the angle of the disk's semi-major axis measured east of north. We set the semi-major axis to be the one located at $90\degc$ counterclockwise from the semi-minor axis associated with the disk near side. The disk near side is the side closest to the central star due to the disk flaring. All distances are taken from Gaia DR3 \citep{gaia_collaboration_vizier_2020}.}
\end{table*}

\newpage
\section{Statistical study using single-size dust particle} 

In this section we suggest estimates for the best dust particles for each SPF category, without applying grain size distribution, and compare this with the case where this distribution is applied (see Fig. \ref{FIG:Statistics}). The first observation is the strong emergence of CAHPs to explain category I SPFs. This dust model becomes the best model, with a monomer size of 200nm, offering a dust size of $\sim2~\mu$m and a relatively high porosity of over 80\%. In this case, and unlike the fractals with a dimension $D_f=1.1$ that dominate when the size distribution is taken into account, amorphous carbon becomes the preferred composition. As for category II, there is also an increase in the number of CAHP considered to be reasonable candidates, but small aggregates with low porosity and irregular grains still dominate, with submicrometric grain size and extremely low porosity, without any major constraints on composition. 

\label{APP:Statistics_single}
\begin{figure}[htbp]
\centering
\includegraphics[width=\textwidth]{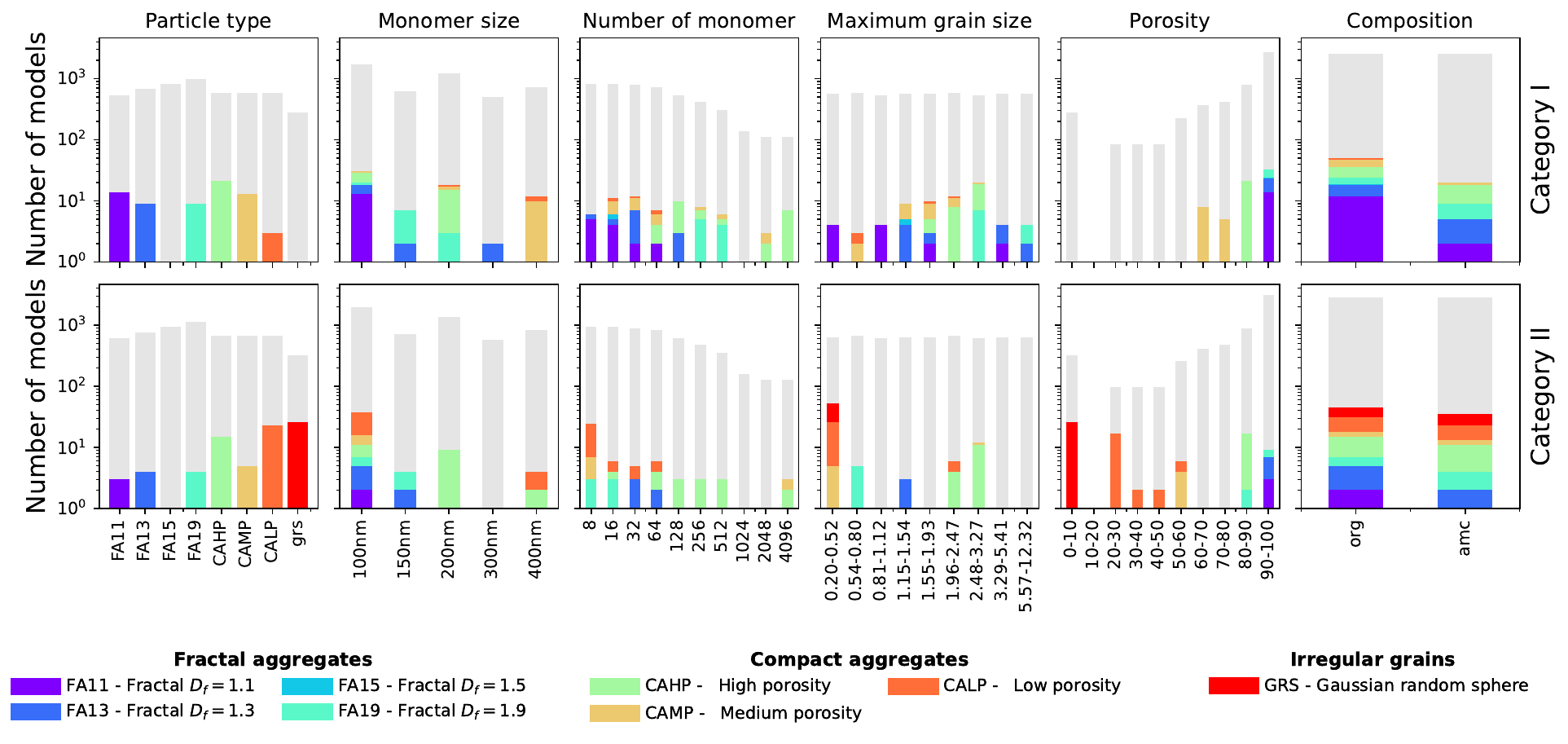}
\caption{Same as Fig. \ref{FIG:Statistics} but without grain size distribution.}
\label{APPFIG:Statistics}
\end{figure}

\newpage
\section{Detailed SPF for all observations} \label{APP:allSPF}
In this section we present the polarized SPF extracted for each system, including when considering each side of the disk separately. Figure \ref{APPFIG:Panel_1} corresponds to the datasets.

\begin{figure}[htbp]
\centering
\includegraphics[width=0.49\textwidth]{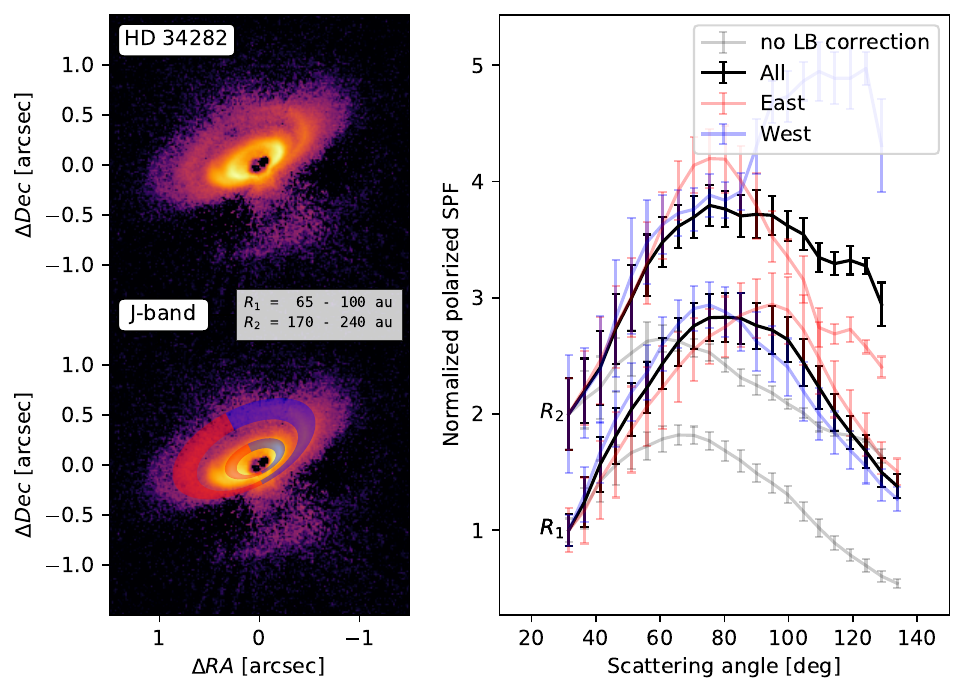}
\includegraphics[width=0.49\textwidth]{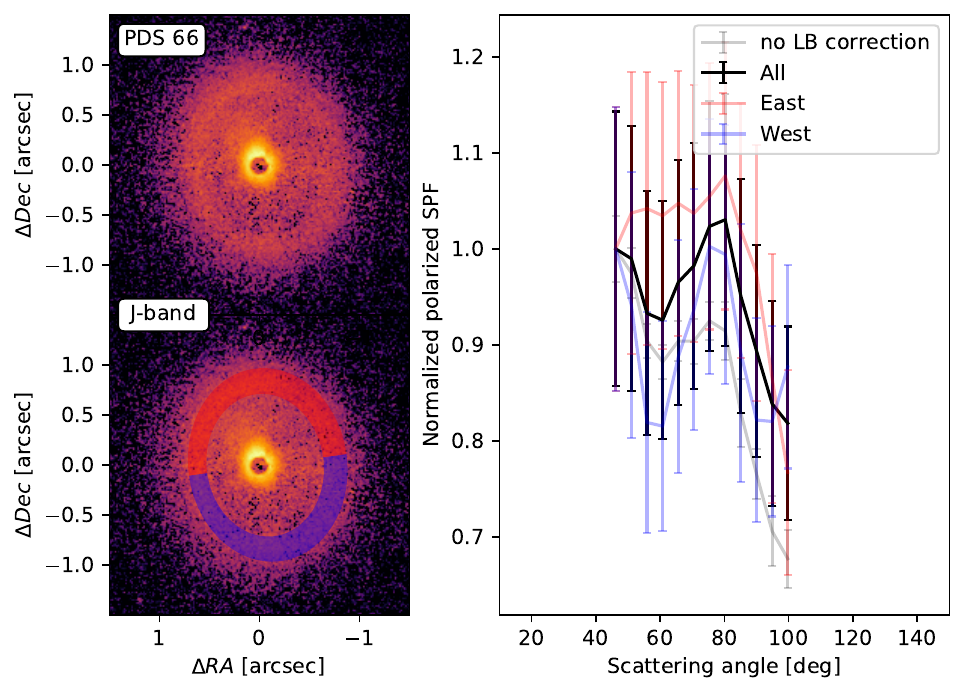}\\[5mm]
\includegraphics[width=0.49\textwidth]{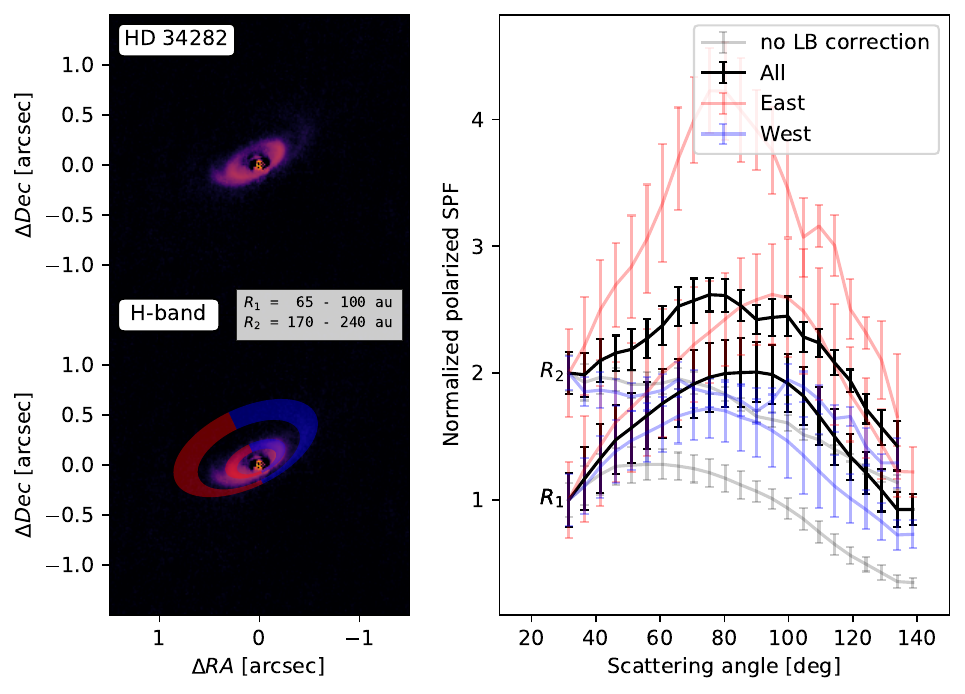}
\includegraphics[width=0.49\textwidth]{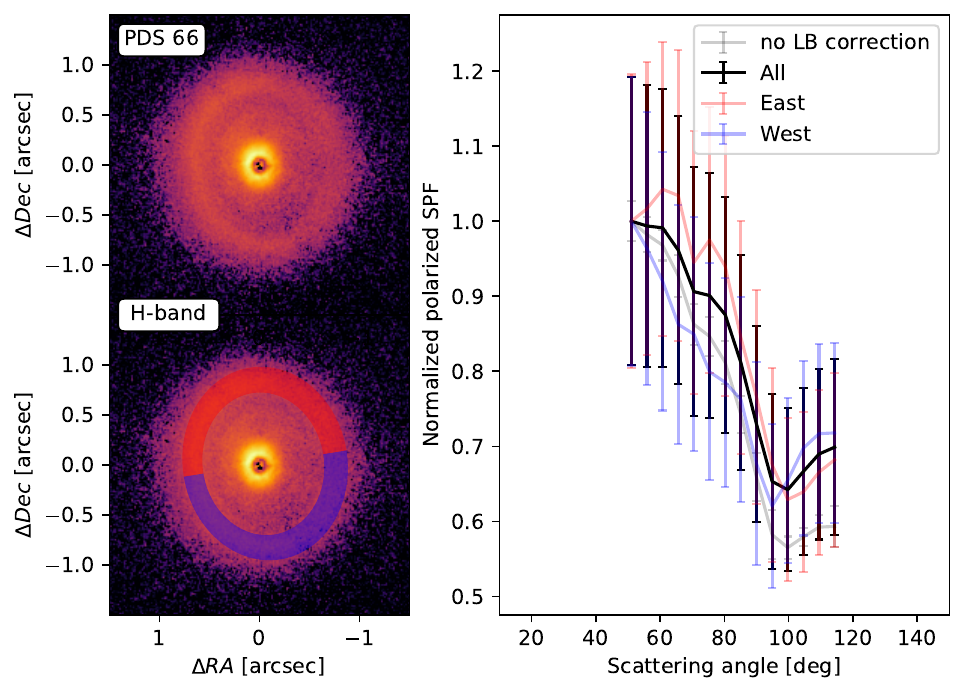}\\[5mm]
\includegraphics[width=0.49\textwidth]{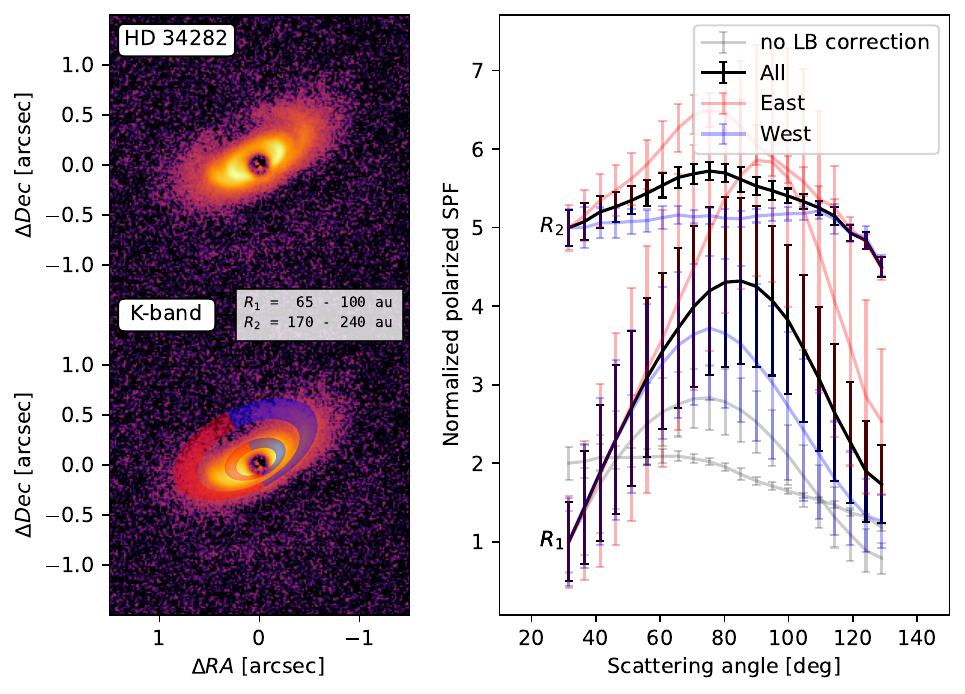}
\includegraphics[width=0.49\textwidth]{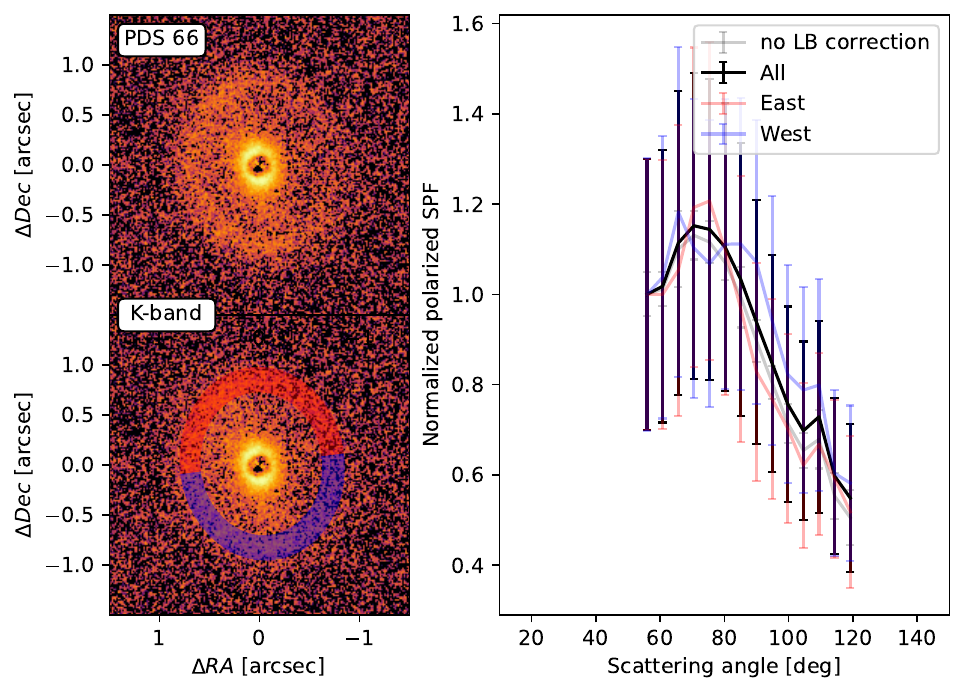}
\caption{Panel of SPF extraction from each side of the ring. In each case, we display the $Q_\phi$ image with and without the extraction zone on the left, and the extracted polarized SPF on the right. We plot SPF for the entire disk (solid black), east side (dotted red), west side (dotted blue), and without correcting for the limb-brightening effect (dotted gray).}
\label{APPFIG:Panel_1}
\end{figure}

\begin{figure}[htbp]
\addtocounter{figure}{-1}
\includegraphics[width=0.49\textwidth]{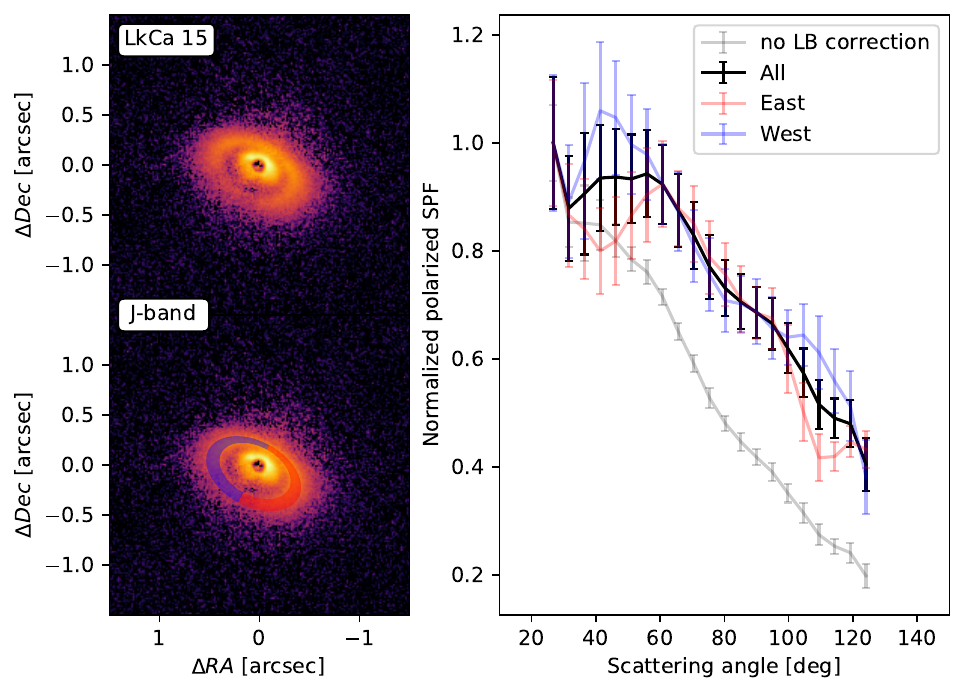}
\includegraphics[width=0.49\textwidth]{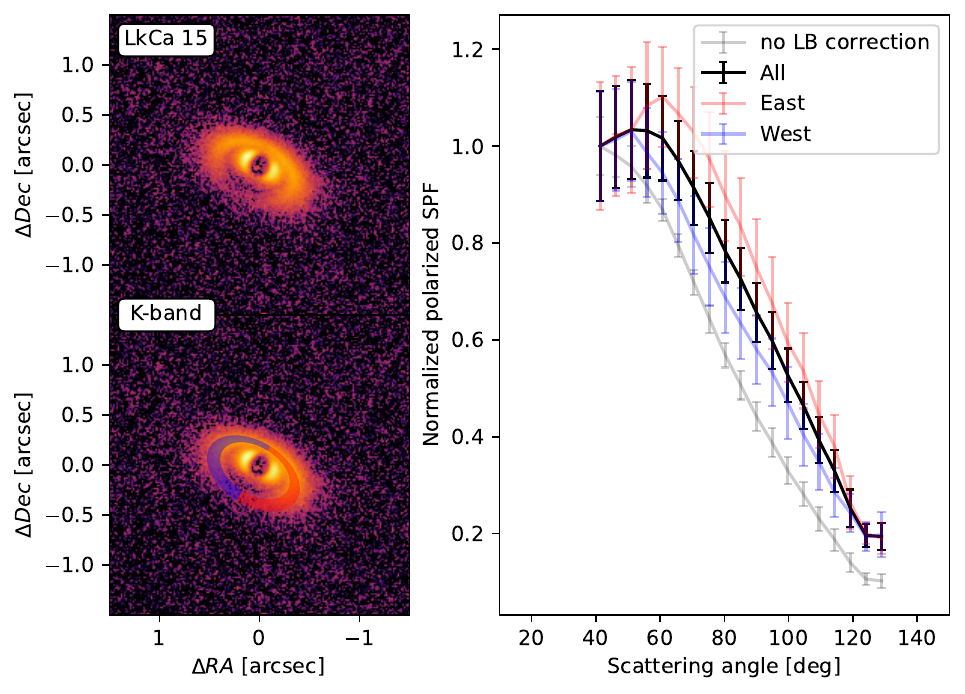}\\[5mm]
\includegraphics[width=0.49\textwidth]{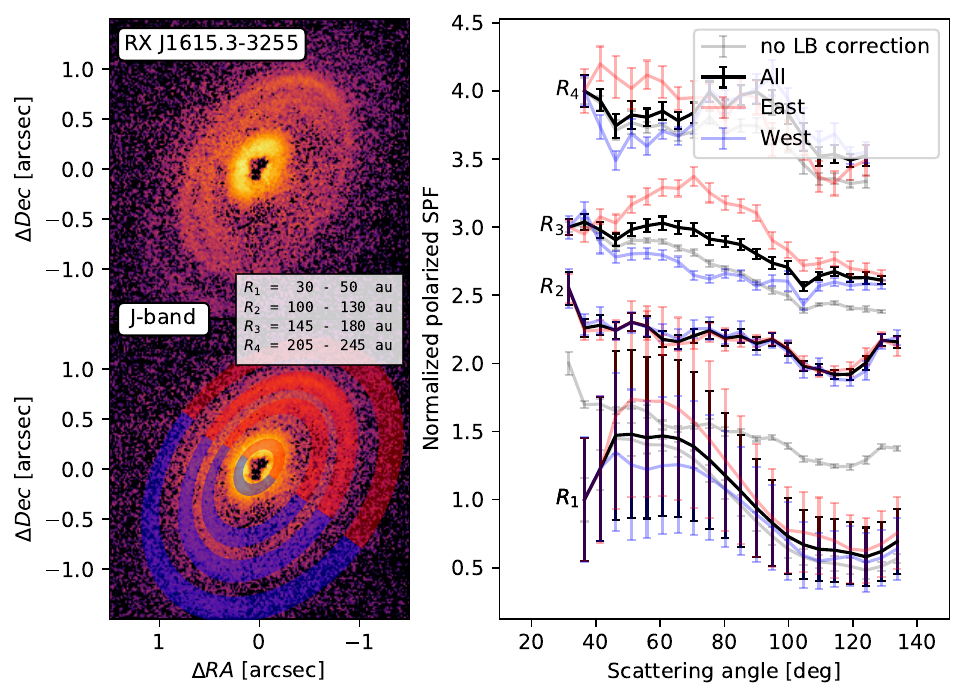}
\includegraphics[width=0.49\textwidth]{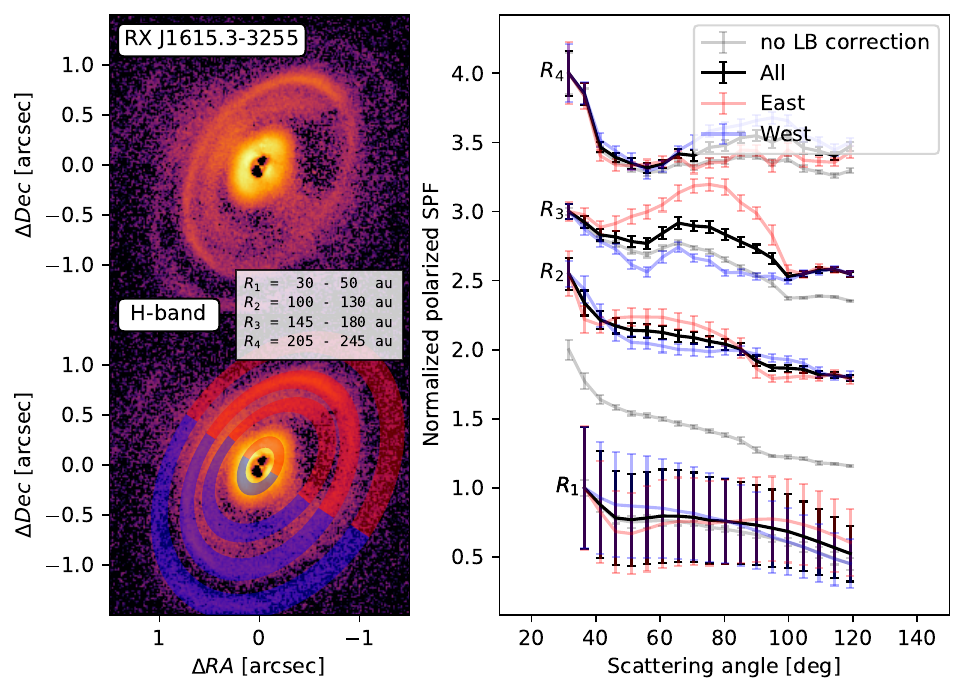}\\[5mm]
\includegraphics[width=0.49\textwidth]{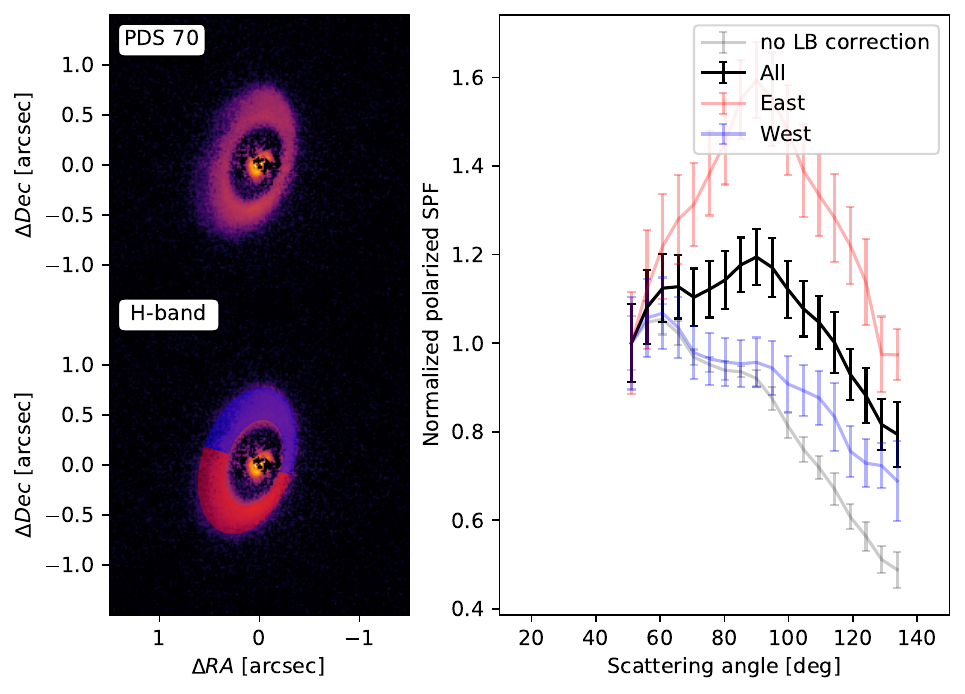}
\includegraphics[width=0.49\textwidth]{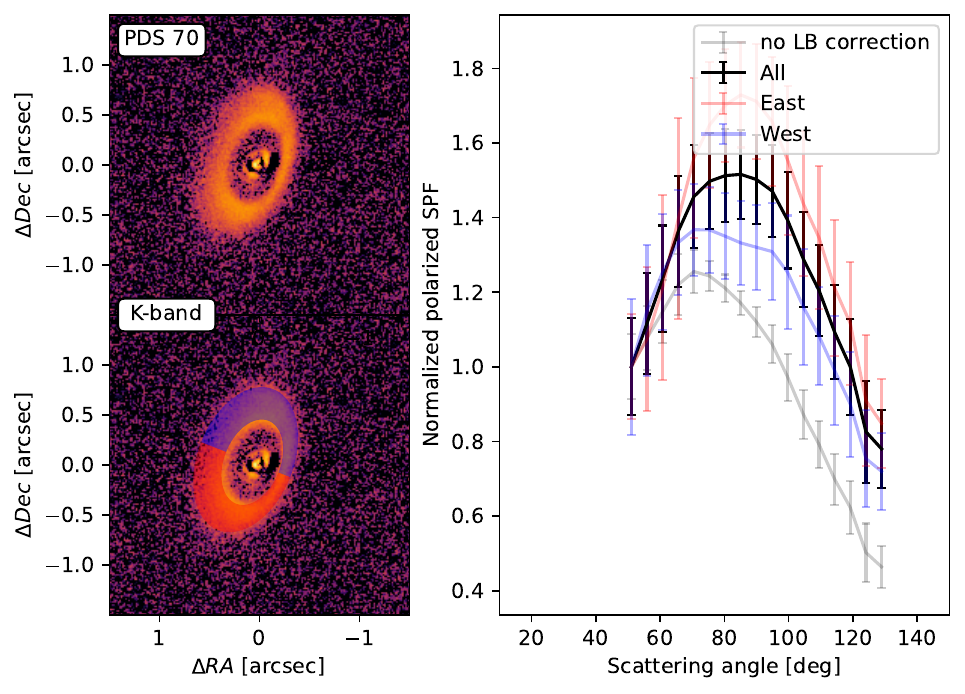}
\caption{Continued.}
\label{APPFIG:Panel_2}
\end{figure}

\begin{figure}[htbp]
\addtocounter{figure}{-1}
\includegraphics[width=0.49\textwidth]{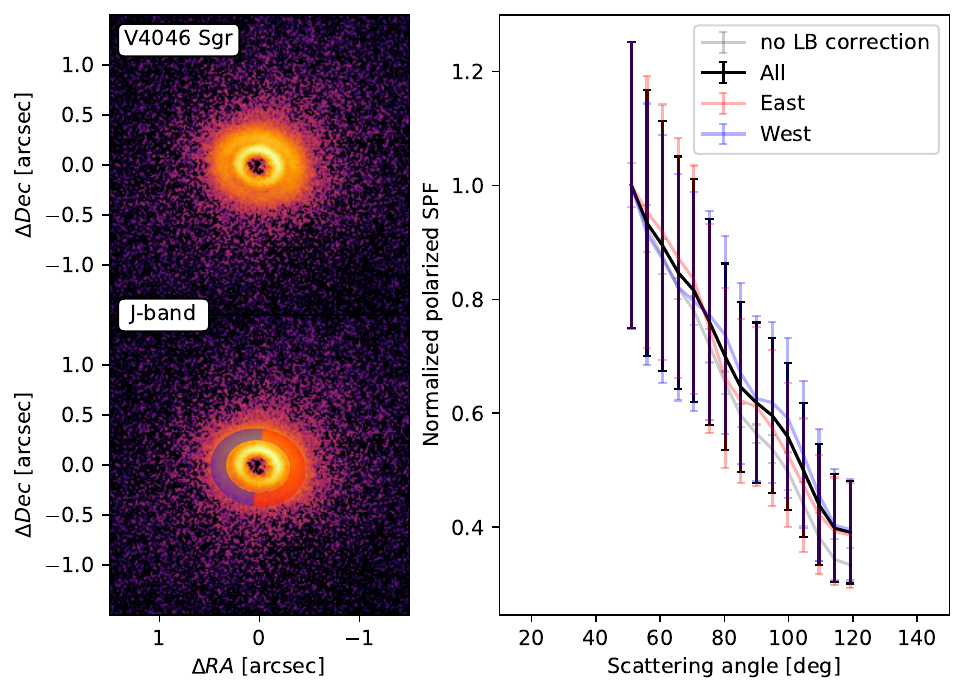}
\includegraphics[width=0.49\textwidth]{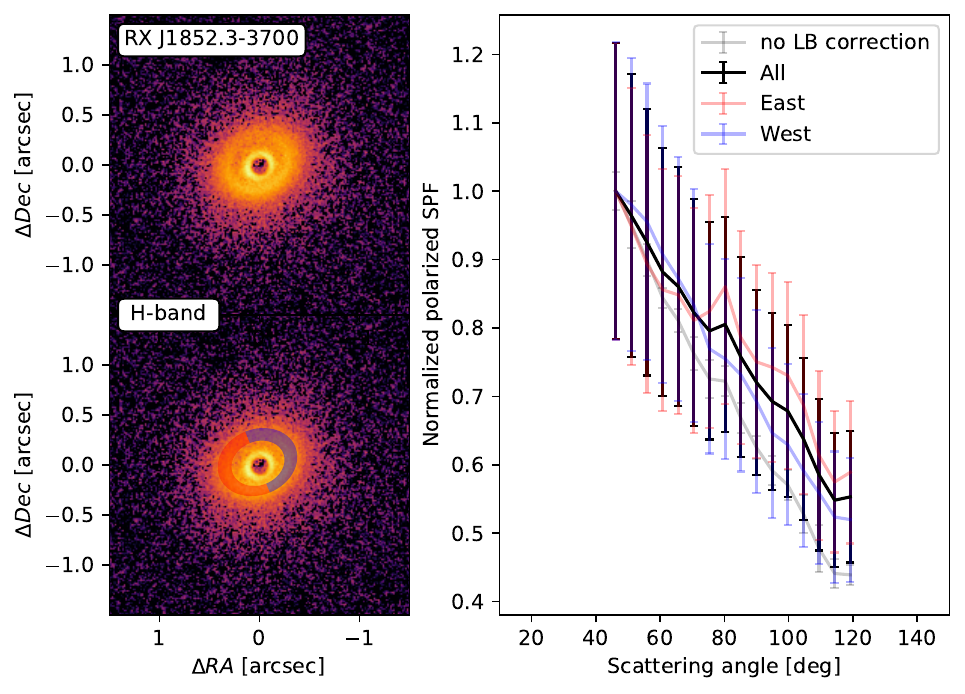}\\[5mm]
\includegraphics[width=0.49\textwidth]{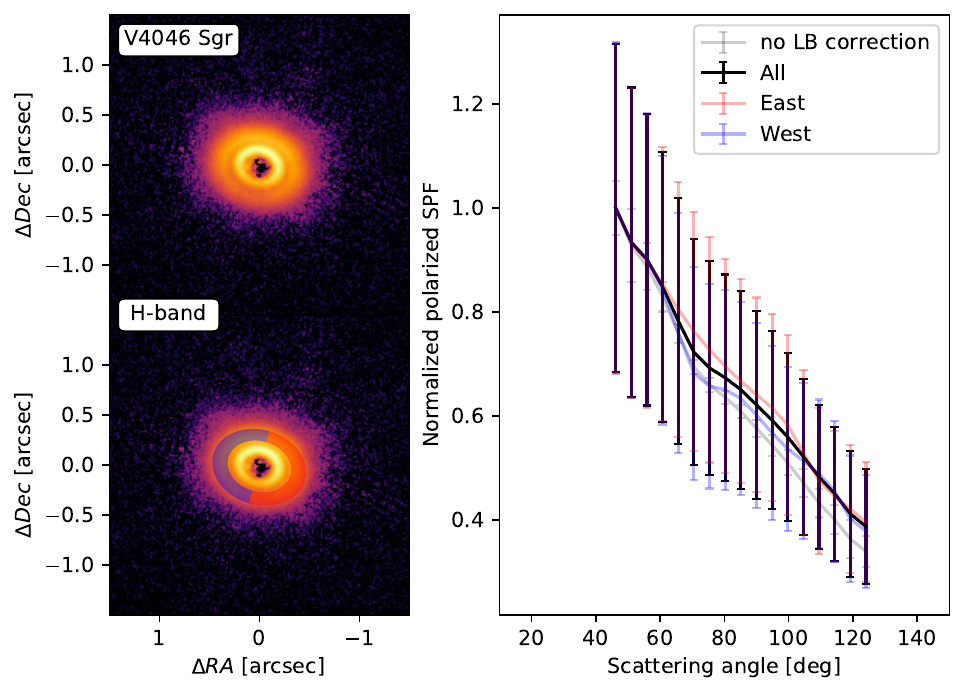}
\includegraphics[width=0.49\textwidth]{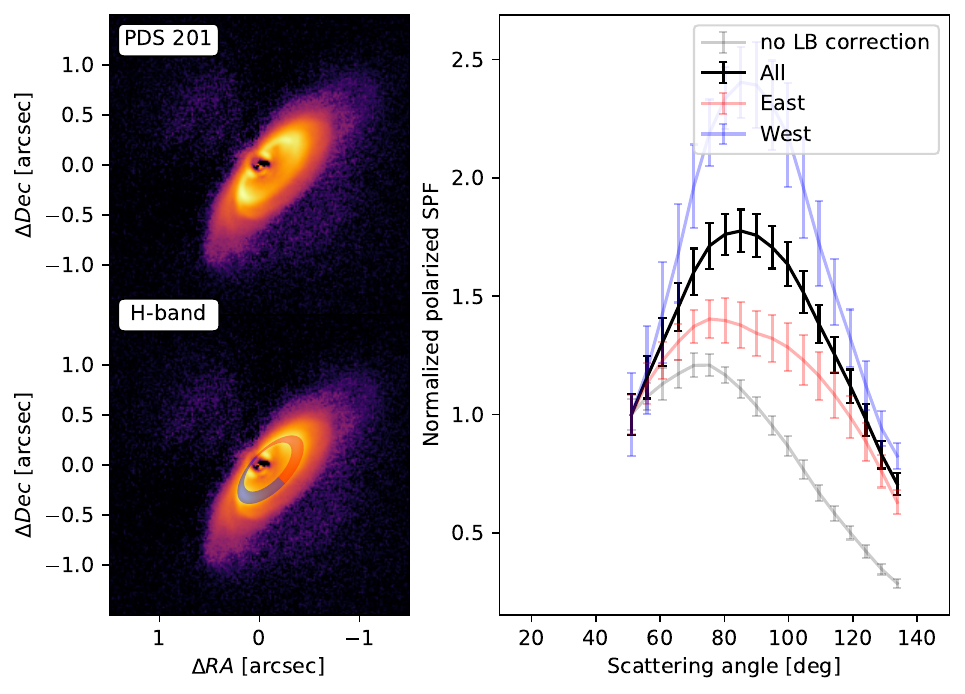}\\[5mm]
\includegraphics[width=0.49\textwidth]{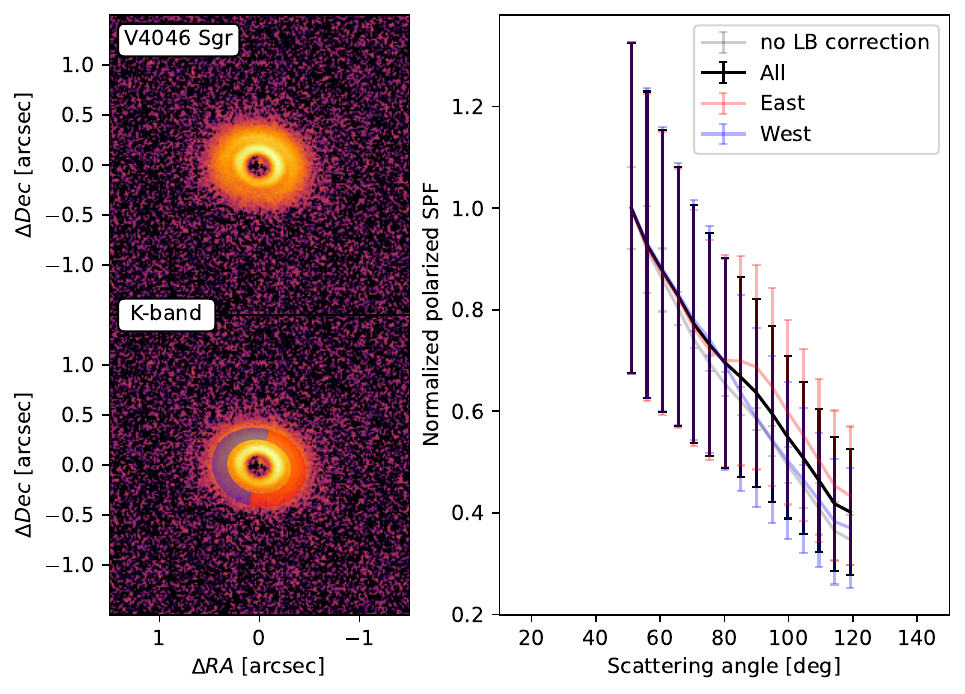}
\includegraphics[width=0.49\textwidth]{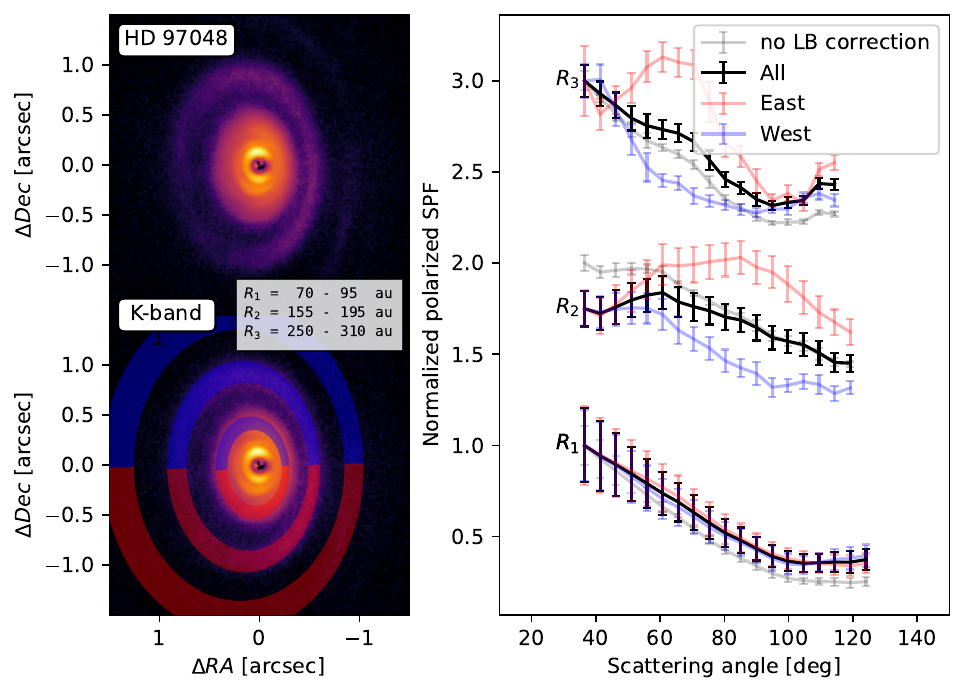}
\caption{Continued.}
\label{APPFIG:Panel_3}
\end{figure}

\begin{figure}[htbp]
\addtocounter{figure}{-1}
\includegraphics[width=0.49\textwidth]{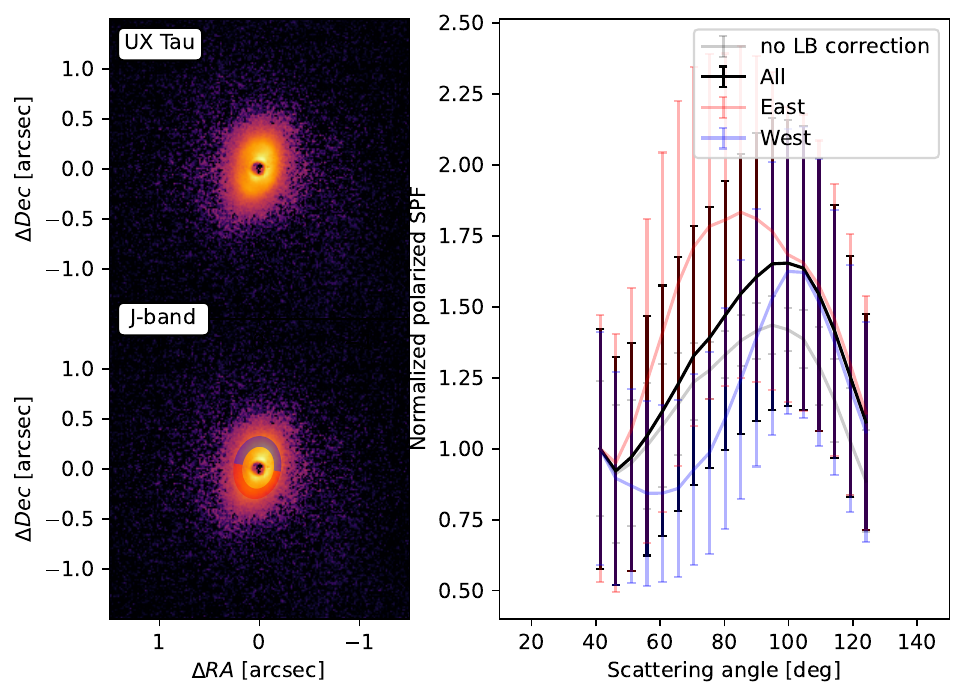}
\includegraphics[width=0.49\textwidth]{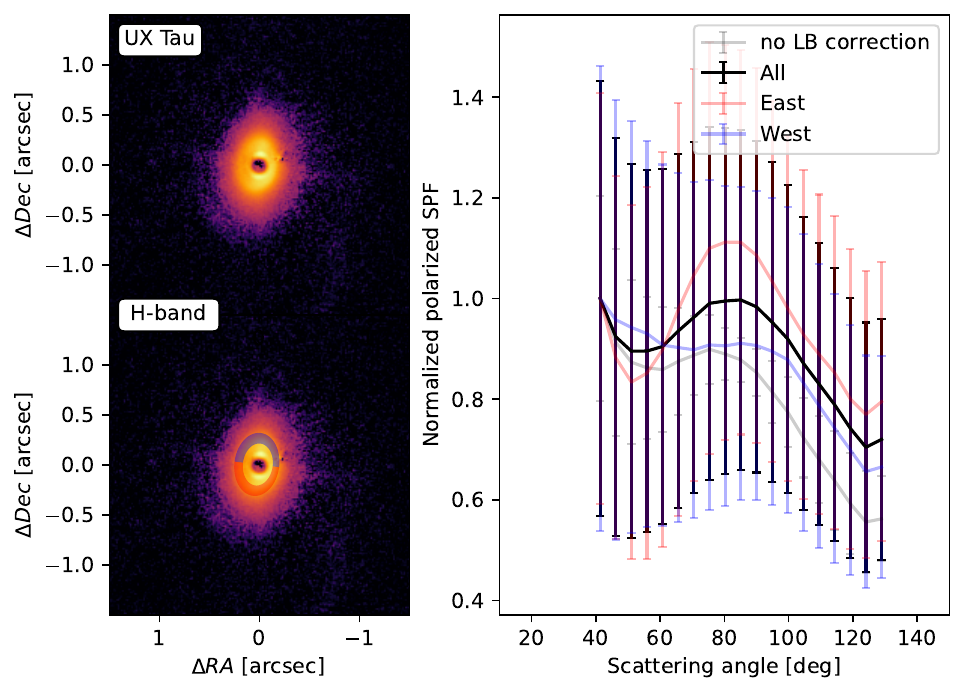}\\[5mm]
\includegraphics[width=0.49\textwidth]{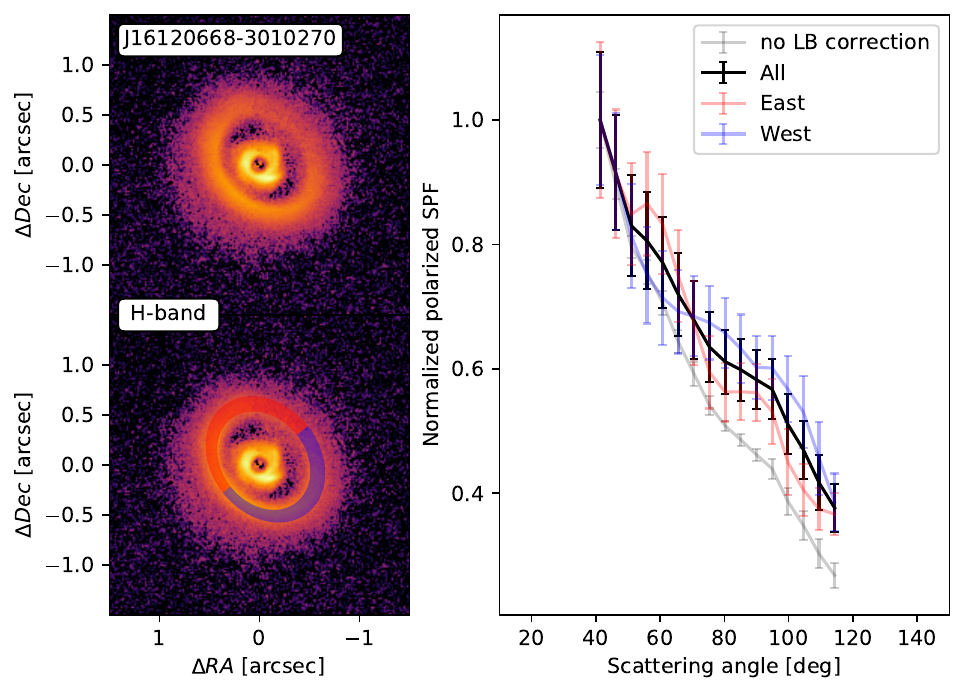}
\includegraphics[width=0.49\textwidth]{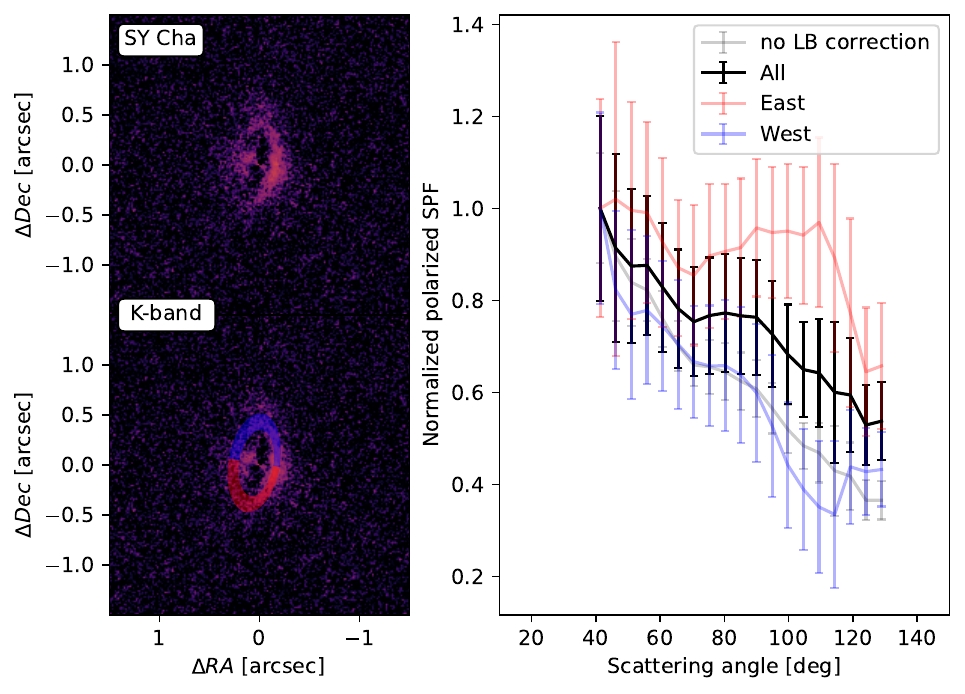}\\[5mm]
\includegraphics[width=0.49\textwidth]{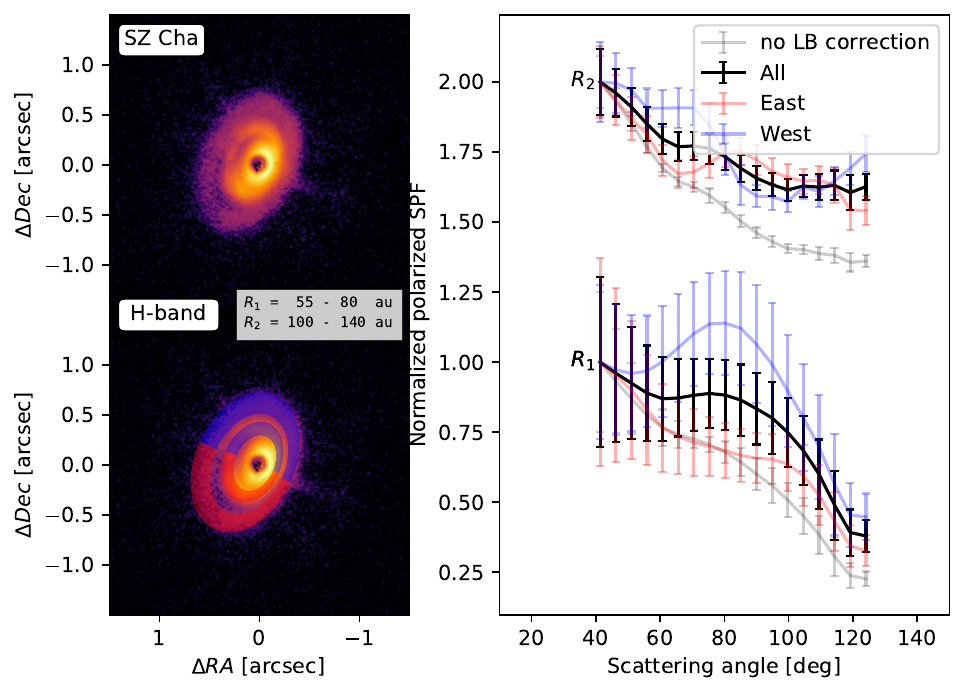}
\includegraphics[width=0.49\textwidth]{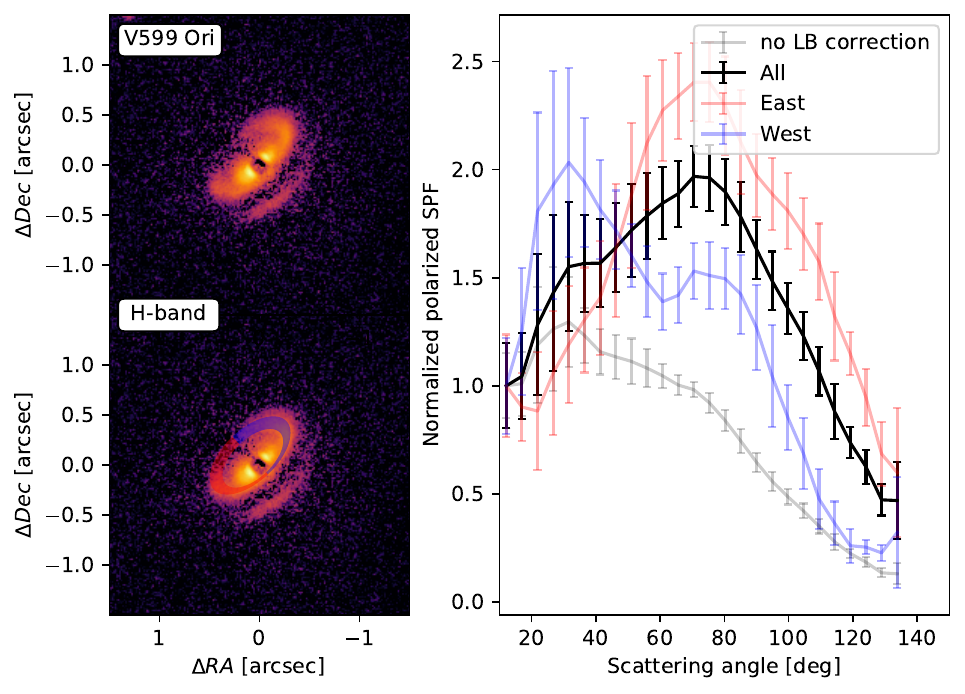}
\caption{Continued.}
\label{APPFIG:Panel_4}
\end{figure}

\begin{figure}[htbp]
\addtocounter{figure}{-1}
\includegraphics[width=0.49\textwidth]{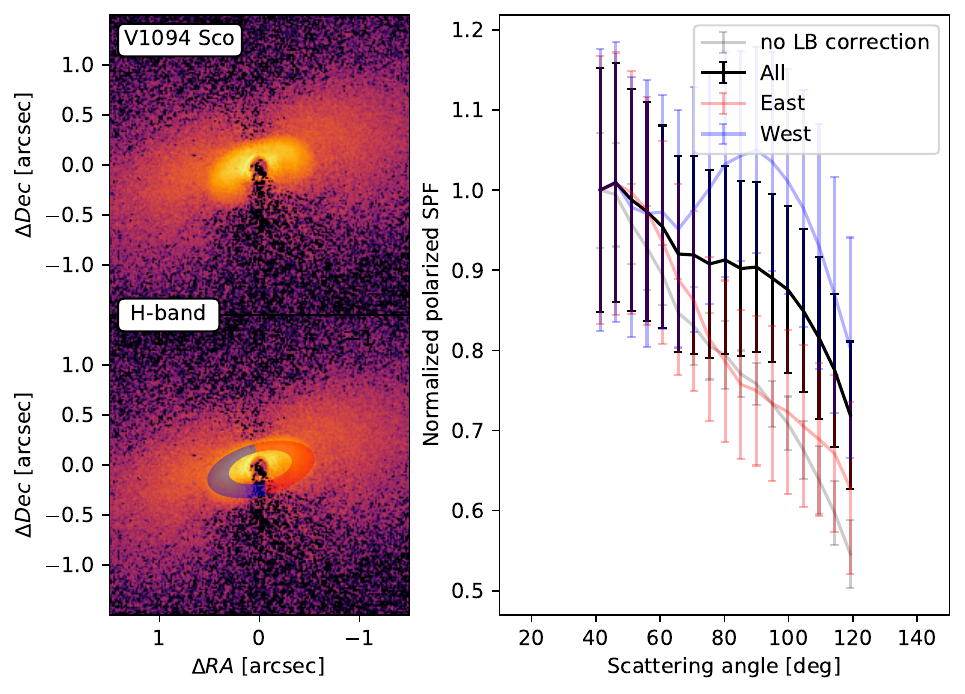}
\includegraphics[width=0.49\textwidth]{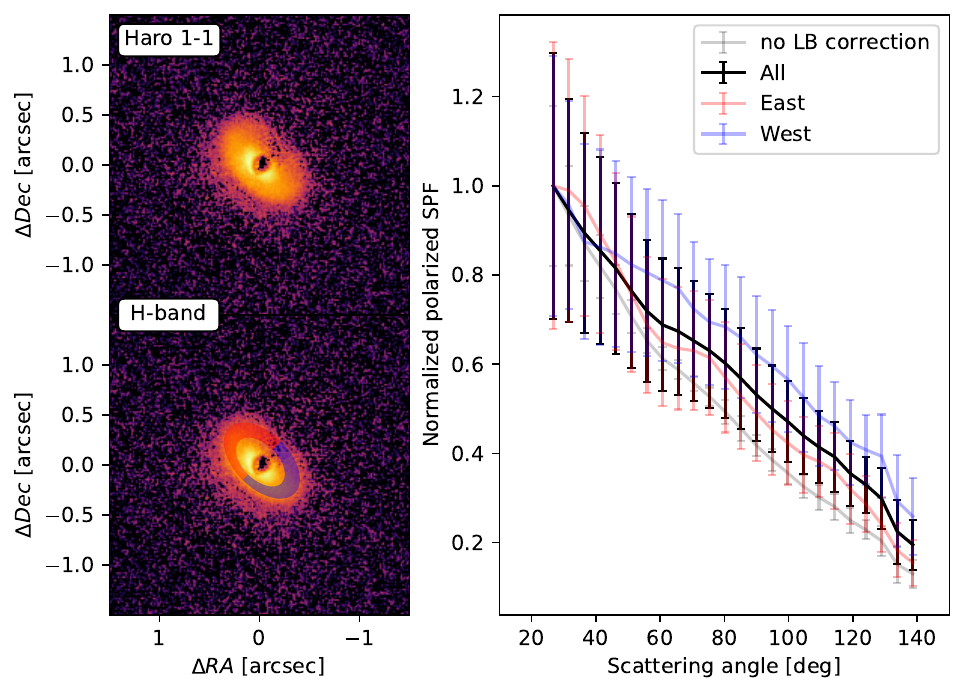}\\[5mm]
\includegraphics[width=0.49\textwidth]{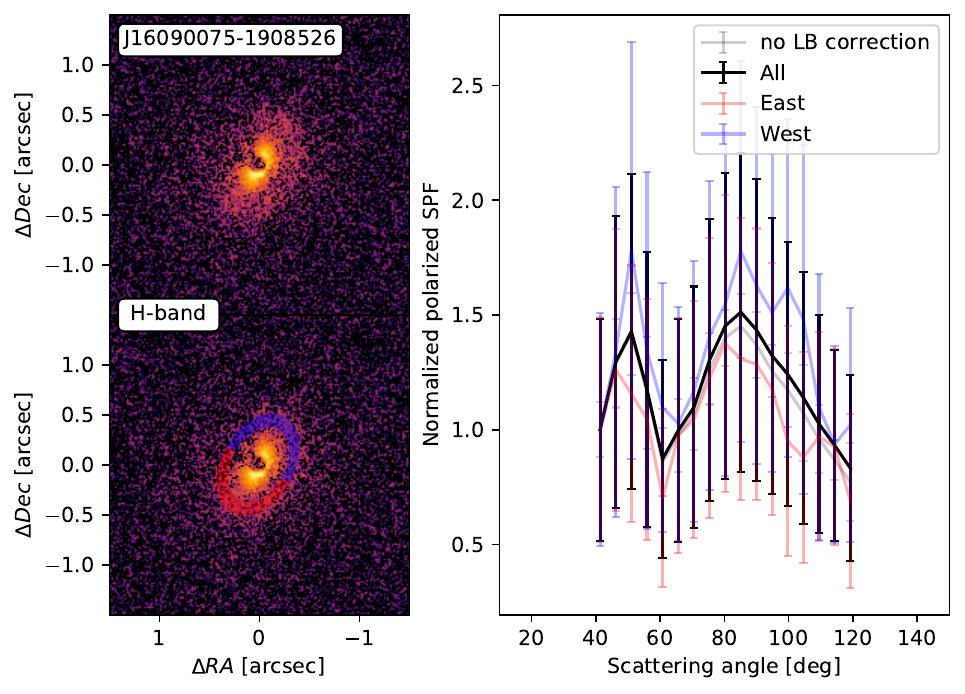}
\includegraphics[width=0.49\textwidth]{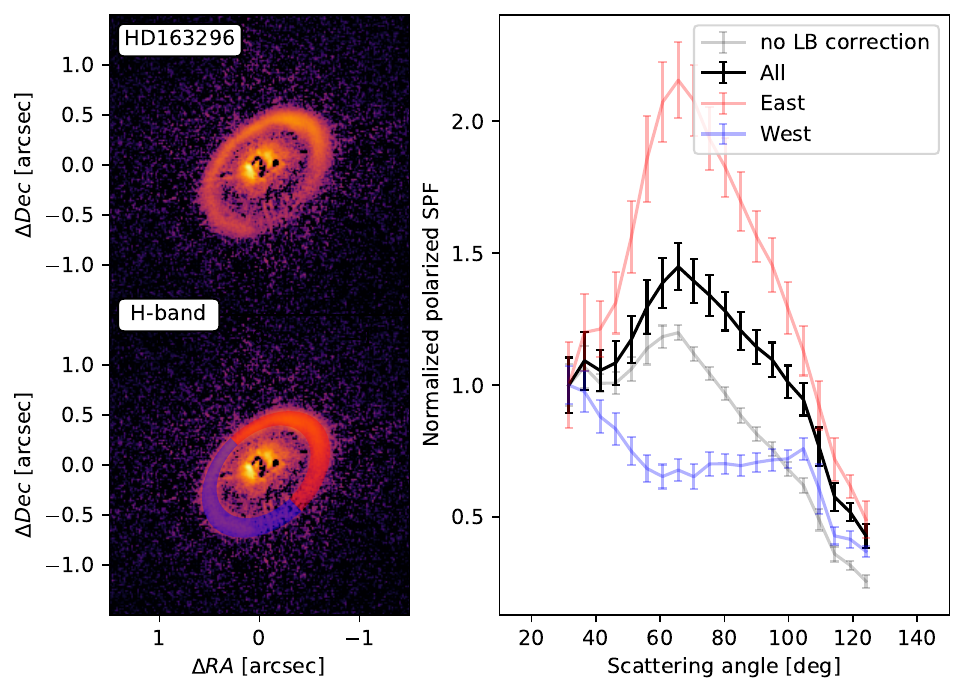}\\[5mm]
\includegraphics[width=0.49\textwidth]{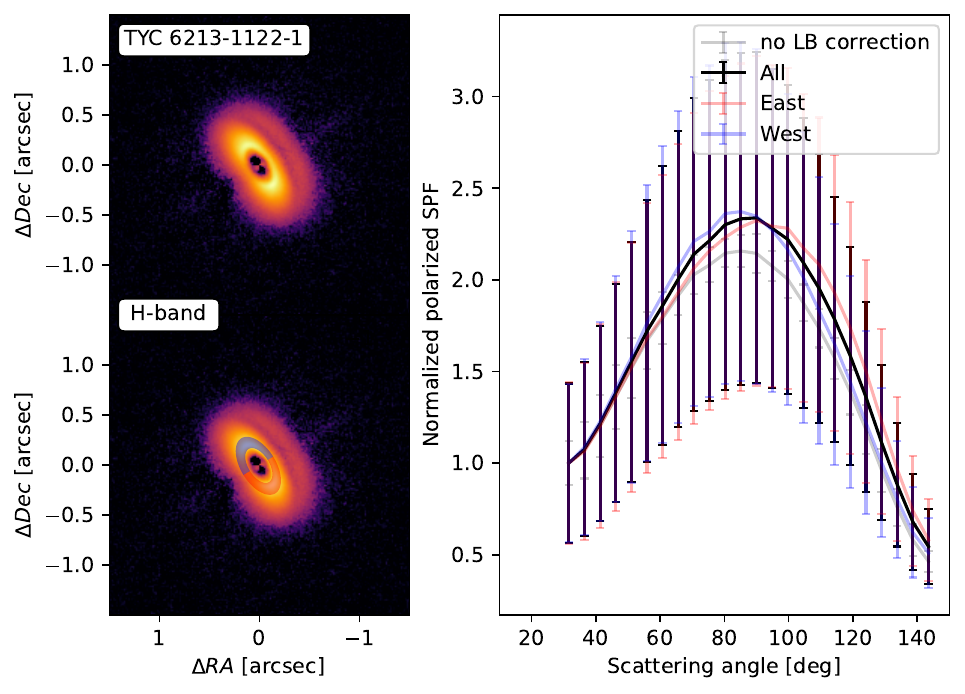}
\includegraphics[width=0.49\textwidth]{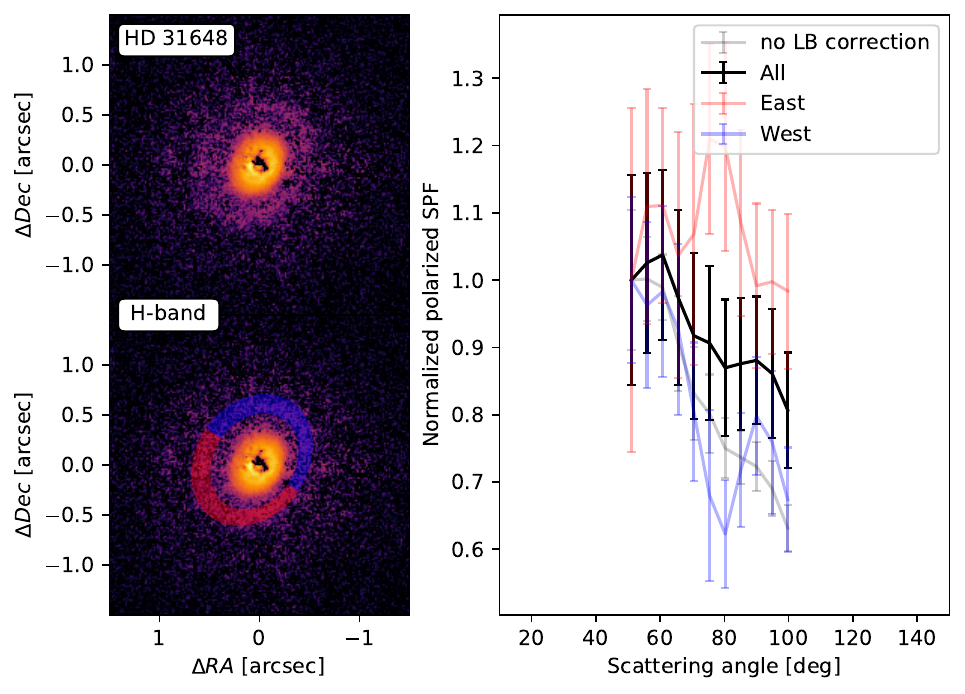}
\caption{Continued.}
\label{APPFIG:Panel_5}
\end{figure}

\end{document}